\documentclass[aps,twocolumn,prc,floatfix,preprintnumbers,superscriptaddress,notitlepage,nofootinbib]{revtex4-1}
\usepackage{dcolumn}
\usepackage{bm}
\usepackage{color}
\usepackage{amssymb}
\usepackage{amsmath}
\usepackage{graphicx}
\usepackage{amsfonts}
\usepackage{slashed}
\usepackage{pstricks}
\usepackage{float}
\usepackage[colorlinks = true,
            linkcolor = blue,
            urlcolor  = blue,
            citecolor = blue,
            anchorcolor = blue]{hyperref}
\usepackage{array}
\allowdisplaybreaks
\usepackage{dcolumn}
\usepackage{epsf}
\usepackage{slashed}

\usepackage{color}
\definecolor{gray}{rgb}{0.6,0.6,0.6}
\definecolor{darkgreen}{rgb}{0.0, 0.545098, 0.0}
\definecolor{darkblue}{rgb}{0.0, 0.0, 0.545098}

\usepackage{color}
\definecolor{BrickRed}{rgb}{0.8, 0.25, 0.33}
\definecolor{gray}{rgb}{0.6,0.6,0.6}
\definecolor{darkgreen}{rgb}{0.0, 0.545098, 0.0}
\definecolor{mypink1}{rgb}{0.858, 0.188, 0.478}

\begin{document}
\preprint{FERMILAB-PUB-23-458-CSAID-ND-T}
\title{One and Two-Body Current Contributions to Lepton-Nucleus Scattering}

\author{Alessandro Lovato}
\affiliation{Physics Division, Argonne National Laboratory, Argonne, Illinois 60439, USA}
\affiliation{INFN-TIFPA Trento Institute of Fundamental Physics and Applications, Via Sommarive, 14, 38123 Trento, Italy}
\author{Noemi Rocco}
\affiliation{Theoretical Physics Department, Fermi National Accelerator Laboratory, P.O. Box 500, Batavia, IL 60410, USA}
\author{Noah Steinberg}
\affiliation{Theoretical Physics Department, Fermi National Accelerator Laboratory, P.O. Box 500, Batavia, IL 60410, USA}

\begin{abstract}
Modeling lepton-nucleus scattering with the accuracy required to extract neutrino-oscillation parameters from long- and short-baseline experiments necessitates retaining most quantum-mechanical effects. One such effect is the interference between one- and two-body current operators in the transition currents, which has been known to enhance the cross-sections, especially in transverse kinematics. In this work, we incorporate such interference in the spectral function formalism, which combines relativistic currents and kinematics with an accurate description of the initial target state. Our analysis of lepton-scattering off $^{12}$C demonstrates that interference effects appreciably enhance the transverse electromagnetic response functions and the flux-folded neutrino-nucleus cross section, in both cases, improving the agreement with experimental data. We discuss the impact on the neutrino-oscillation program and the determination of nucleon axial form factors.

\end{abstract}

\maketitle

\section{Introduction}\label{sec:intro}
Neutrino-oscillation experiments rely on precise theoretical calculations of neutrino interactions with target nuclei to accurately extract oscillation parameters. These calculations will become even more relevant for high-statistics DUNE and Hyper-Kamiokande~\cite{DUNE:2015lol,T2K:2021xwb,Hyper-Kamiokande:2018ofw}, where modeling neutrino-nucleus cross sections constitutes a primary source of systematic uncertainty~\cite{Benhar:2015wva,Katori:2016yel,NuSTEC:2017hzk,Ruso:2022qes}. While inclusive cross-sections are crucial for determining the energy distribution of charged-current events based on the kinematics of the outgoing charged lepton, models capable of providing accurate semi-exclusive predictions for the final state become invaluable for correct calorimetric energy reconstruction~\cite{Ankowski:2015jya}. The vast amount of available electron scattering data, combined with the increasing body of measurements from the accelerator neutrino program, has highlighted deficiencies in models of lepton-nucleus cross sections routinely used in neutrino event generators~\cite{MINERvA:2019gsf,T2K:2020jav,CLAS:2021neh,MINERvA:2019ope,MINERvA:2021wjs,T2K:2021naz,MicroBooNE:2023cmw,MicroBooNE:2022emb,MINERvA:2022mnw,MicroBooNE:2023tzj,MINERvA:2023kuz}. These limitations pose a major obstacle to achieving the precision design goal of DUNE, which requires percent-level accuracy in neutrino-nucleus cross sections~\cite{DUNE:2020fgq}.

This challenge has served as a catalyst for improving theoretical models of neutrino-nucleus scattering. Accurate predictions at moderate energies and momentum transfers have been achieved using ``ab-initio'' methods, starting from consistent Hamiltonians and currents. Continuum Quantum Monte Carlo~\cite{Carlson:2014vla,Gandolfi:2020pbj} (QMC) and, specifically, the Green's Function Monte Carlo (GFMC) methods have been successfully employed to compute electroweak response functions of nuclei up to $^{12}$C in the quasielastic (QE) region~\cite{Lovato:2020kba,Lovato:2017cux,Lovato:2016gkq}. These calculations retain quantum many-body correlations in both the initial and final states of the reaction that are consistent with one- and two-body current operators. More recently, Coupled Cluster (CC) theory has been applied to carry out the first ab-intio calculation of the electromagnetic response functions of $^{40}$Ca \cite{Sobczyk:2023sxh}, including one-body currents only. Despite this remarkable accomplishment, due to the slow convergence of single-particle basis expansion, CC must use relatively soft nuclear forces as input~\cite{Ekstrom:2015rta,Jiang:2020the}, limiting its applicability in modeling the final state of the reaction to momentum transfers $q\lesssim 400$ MeV. In addition, both the GFMC and CC are constrained by their non-relativistic nature, although some relativistic effects can be controlled by performing calculations in a specific reference frame~\cite{Rocco:2018tes,Nikolakopoulos:2023zse}. Most importantly, they can only predict inclusive observables, while  liquid argon time projection chambers have excellent calorimetric reconstruction~\cite{MicroBooNE:2021ddy}, which is enabled by its unique combination of size, position resolution precision, and low energy thresholds~\cite{Castiglioni:2020tsu}.  

These novel detector capabilities and the high energies characterizing most neutrino beams have motivated the development of theoretical models that can accommodate fully-relativistic kinematics and currents and provide detailed information on the kinematic variables associated with the hadronic final states. A key role in this regard is played by methods relying on the factorization of the final state of the reaction, such as the spectral-function (SF) formalism~\cite{Benhar:1993ja,Benhar:2006wy,Rocco:2020jlx} and the short-time approximation (STA)~\cite{Pastore:2019urn}, which retain the vast majority of correlation effects in both the initial and final states of the reaction. Both the recently introduced QMC-SF~\cite{CLAS:2022odn}, constructed from spectroscopic overlaps and two-nucleon momentum distributions computed within the variational Monte Carlo method, and the STA adopt the same model of nuclear dynamics as the GFMC for describing the initial target state. Therefore, benchmarking the corresponding predictions for the inclusive electron-nucleus scattering cross-section, as in Ref.~\cite{Andreoli:2021cxo} for $A=3$ nuclei, brings to light uncertainties inherent in factorization schemes and relativistic effects. 

A recent comparison on electron- and neutrino scattering off $^{12}$C between the GFMC and QMC-SF~\cite{Nikolakopoulos:2023zse} has demonstrated the accuracy of the GFMC at intermediate momentum transfer, provided that GFMC calculations are performed in a reference frame that minimizes nucleon momenta and the kinematics of the reaction is corrected leveraging the ``two-fragment'' model~\cite{Efros:2009qp,Rocco:2018tes}. Assessing the uncertainty inherent in the factorization of the final state when two-body currents are present is hindered by the fact that GFMC includes interference between one- and two-body currents, generally neglected in the SF formalism. Refs.~\cite{Benhar:2015ula,Rocco:2015cil} are notable exceptions; there the interference leading to two-particle two-hole final states was considered. However, it was found to peak in the so-called ``dip'' region, while in GFMC interference effects enhance the transverse response function in the QE region~\cite{Lovato:2020kba,Lovato:2017cux,Lovato:2016gkq}.  

Consistent with the GFMC, both within the STA and relativistic mean-field approaches, the interference between one- and two-body contributions has been found to enhance the transverse electromagnetic response function in the QE region~\cite{Pastore:2019urn, Franco-Munoz:2022jcl, Franco-Munoz:2023zoa}. A similar pattern emerges from correlated-basis functions calculations of infinite nuclear matter, and it has been ascribed to tensor correlations in the nuclear many-body wave function~\cite{Fabrocini:1996bu}. On the other hand, relativistic Fermi Gas calculations report either similar constructive interference in the transverse channel~\cite{Umino:1995bql} or negligible effects~\cite{Amaro:2002mj}. In this work, we significantly improve the accuracy of the QMC-SF formalism by including one-nucleon knockout processes induced by the interference between one- and two-body currents. The relative importance of these processes with respect to pure one-body and two-body contributions is analyzed in detail.

This work is organized as follows. In Sec.~\ref{sec:currents}, we introduce the SF formalism and the factorization scheme, reviewing the calculation of the ``diagonal'' one-body one-body and two-body two-body current contributions. The extension of this formalism to accommodate the interference term between one- and two-body currents is discussed in Sec.~\ref{sec:interf}, while the explicit expression of the two-body current operator is reported in Sec.~\ref{sec:currents}. In Sec.~\ref{sec:results}, we show the improved agreement between theoretical calculations that account for the interference contributions and both electron-nucleus and neutrino-nucleus scattering data. Finally, in Sec.~\ref{sec:conclusion}, we draw our conclusions and outline future perspectives of this work.

\section{Factorization Scheme with One- and Two-Body Currents}\label{sec:currents}
In the one-boson exchange approximation, the inclusive lepton-nucleus scattering cross section can be expressed in terms of the response tensor, which contains all information on nuclear dynamics
\begin{equation}\label{eq:ResponseFunctionE}
    R^{\mu\nu} = \sum_{f}\langle 0|J^{\mu\dagger}|f\rangle\langle f|J^{\nu}|0\rangle\delta (E_{0} + \omega - E_{f})\,.
\end{equation}
In the above equation, $|0\rangle$ is the nuclear ground state, $|f\rangle$ denotes the hadronic final state, and $J^{\mu}$ is the electroweak current operator. While factorization schemes can accommodate real-pion production~\cite{Rocco:2019gfb}, in this work we restrict our analysis to final states containing nucleons only. The latter are the most relevant ones for lower-energy experiments, like T2K~\cite{T2K:2023smv} and the SBN program~\cite{Machado:2019oxb}, whose neutrino fluxes peak at or below $1$ GeV. In this regime, the nuclear current operator consists of a sum of one- and two-body contributions
\begin{equation}
    J^{\mu} = \sum_{i}j^{\mu}_{i} + \sum_{j>i}j^{\mu}_{ij}\, ,
\end{equation}
In the QE region, the dominant reaction mechanism is single nucleon knockout. For large values of momentum transfer, the hadronic final state can be approximated by
\begin{equation}
    |f\rangle \simeq  |p'\rangle \otimes |f_{A-1}\rangle,
    \label{eq:fact:1b}
\end{equation}
where $|p'\rangle$ is the plane-wave state describing the nucleon  produced at the interaction vertex, and $|f_{A-1}\rangle$ is the spectator system, carrying momentum $\textbf{p}_{A-1}$.

Inserting this factorization ansatz as well as a single-nucleon completeness relation leads to the following expression for the one-body matrix element
\begin{equation}\label{eq:factorize1b}
    \langle 0|J^{\mu}|f\rangle \rightarrow \sum_{h}\langle 0|[|h\rangle\otimes|f_{A-1}\rangle]\langle h|\sum_{i}j^{\mu}_{i}|p'\rangle\, .
\end{equation}
The first term in the above summation can safely be computed within nonrelativistic nuclear many body theory, as it is independent of momentum transfer. The second piece is the nucleon level matrix element, which is fully specified by the relativistic one-body currents and single-particle plane wave states. The incoherent one-body current contribution to the response tensor can then be expressed as
\begin{equation}\label{eq:W1body}
\begin{aligned}
    R_{1b}^{\mu\nu}(\textbf{q},\omega) = &\int\frac{d^{3}h}{(2\pi)^3}dEP_{h}(\textbf{h},E)\frac{m^{2}_{N}}{e(\textbf{h})e(\textbf{h} + \textbf{q})} \\ 
    & \times\sum_{i}\langle h|j_{i}^{\mu\dagger}|h +q\rangle\langle h+q|j_{i}^{\nu}|h\rangle \\ 
    & \times\delta(\tilde{\omega} + e(\textbf{h}) - e(\textbf{h}+\textbf{q})),
\end{aligned}
\end{equation}
where we use the relativistic expression of the single-particle energy $e(\mathbf{k}) = \sqrt{m_N^2 + \mathbf{k}^2}$. The factors $m_{N}/e(\textbf{h})$ and $m_{N}/e(\textbf{h} + \textbf{q})$ are included to account for the covariant normalization of the Dirac spinors in the matrix elements of the relativistic current. The energy transfer is replaced by $\tilde{\omega} = \omega + m_{N} - E - e(\textbf{h})$ to account for the fact that the initial nucleon is bound. The expression for the response tensor given in Eq.~\eqref{eq:W1body} is identical for electron and neutrino-nucleus scattering --- the corresponding one-body current electromagnetic and electroweak operators can be found in Ref.~\cite{Lovato:2023raf}. Note that for the axial form factor we consider both the dipole parameterization with $M_{A} = 1.0$ GeV, as well as recent results from Lattice Quantum Chromo Dynamics (LQCD)~\cite{Park:2021ypf,Djukanovic:2022wru,RQCD:2019jai}. 

The single-nucleon spectral function $P_{h}(\textbf{h},E)$ provides the probability of removing a nucleon with momentum $\textbf{h}$ and leaving the residual nucleus with an excitation energy $E$~\cite{Benhar:2006wy}
\begin{equation}
P_{h}(\textbf{h},E) = \sum_f |[_{A-1}\langle f | \otimes \langle k | ] |0\rangle|^2 \delta(E-E_f^{A-1}+E_0)
\end{equation}
It can be expressed as a sum of two components
\begin{equation}
P_{h}(\textbf{h},E) = P_{\rm MF}(\textbf{h},E) + P_{\rm corr}(\textbf{h},E)\,.
\end{equation}
The mean-field term corresponds to $|f_{A-1}\rangle$ remaining in a bound state, while the correlation contribution involves at least nucleon in the continuum. The one used in this work is based on QMC calculations of spectroscopic overlaps and two-body momentum distributions to obtain $P_{h}^{\rm MF}(\textbf{h},E)$ and $P_{h}^{\rm corr}(\textbf{h},E)$, respectively~\cite{Lovato:2023raf,Simons:2022ltq,CLAS:2022odn}, and it is normalized as
\begin{equation}
\int \frac{d^3 k}{(2\pi)^3}dE P_h(\mathbf{k},E) = 1\, .
\end{equation}

In the results presented, we also include the contribution of two-body currents commonly denoted as meson exchange currents (MEC), whose primary effect is to induce two-nucleon knockout.
This channel is subdominant with respect to single-nucleon knockout in the QE region, but it has been shown to provide important strength at intermediate values of energy transfer between the QE peak and resonance production region ~\cite{Dekker:1994yc,DePace:2003spn,Rocco:2015cil}. Two-nucleon emission processes induced by MEC play a significantly enhance the flux-folded neutrino-scattering cross section, bringing theoretical predictions closer to experimental measurements ~\cite{MINERvA:2015ydy,NOvA:2020rbg,MINERvA:2021wjs,Martini:2010ex,Martini:2011wp,Nieves:2011pp,Nieves:2011yp,Megias:2014qva,Gonzalez-Jimenez:2014eqa,Pandey:2014tza}. 

MEC leading to two-nucleon knockout have been incorporated in the factorization scheme as discussed in Refs.~\cite{Benhar:2015ula,Lovato:2023raf,Rocco:2015cil,Rocco:2018mwt,Simons:2022ltq}. Here, we report the final expression for the pure two-body current component of the response tensor, associated with two-nucleon emission
\begin{align}\label{eq:twonucleon_R}
&R^{\mu\nu}_{\rm 2b}({\bf q},\omega)=\frac{V}{4} \int dE \frac{d^3h}{(2\pi)^3}  \frac{d^3h^\prime}{(2\pi)^3}\frac{d^3p}{(2\pi)^3}
\frac{m_N^4}{e({\bf h})e({\bf h^\prime})e({\bf p})e({\bf p^\prime})} \nonumber \\
 &\qquad \times  P_h^{\rm NM}({\bf h},{\bf h}^\prime,E) \sum_{ij}\, \langle h\, h^\prime | {j_{ij}^\mu}^\dagger |p\,p^\prime\rangle_a \,_a\langle p\,p^\prime |  j_{ij}^\nu | h\, h^\prime \rangle \nonumber \\
 &\qquad \times \delta(\omega-E+2m_N-e(\mathbf{p})-e(\mathbf{p}^\prime))\, .
\end{align}
The normalization volume $V=\rho / A$ with $\rho=3\pi^2 k_F^3/2$ depends on the Fermi momentum of the nucleus, which for $^{12}$C is taken to be $k_F=225$ MeV. The factor $1/4$ accounts for the fact that we sum over indistinguishable tow-particle two-hole final state. The two-nucleon spectral function in Eq.~\eqref{eq:twonucleon_R} is computed using QMC techniques and retains correlations between the two struck particles~\cite{Simons:2022ltq}.

\begin{figure}[h]
    \includegraphics[width=0.4\textwidth]{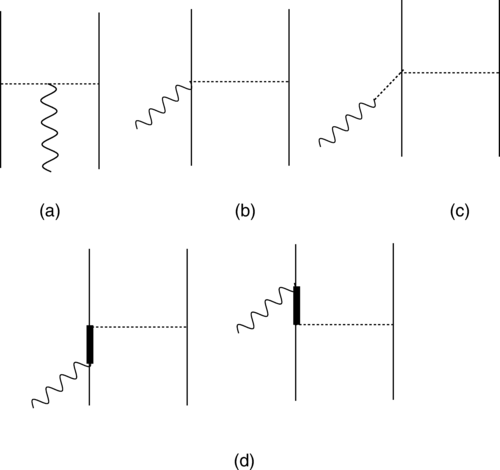} 
    \caption{Feynman diagrams describing two-body currents contributions associated to: pion in flight (a), seagull (b), pion-pole (c), and delta excitations (d) processes. Solid, thick, and dashed lines correspond to nucleons, deltas, pions, respectively. The wavy line represents the vector boson.}
    \label{mec:diag}
\end{figure}

The electroweak MEC used in this work are those of Ref.~\cite{RuizSimo:2016rtu}, which have been derived by coupling to a second nucleon line the pion-production amplitudes obtained within the
non-linear $\sigma$ model in Ref.~\cite{Hernandez:2007qq}. Their electromagnetic counterparts can be obtained by setting the axial pieces of each current to zero and modifying their isospin dependence. The MEC is the sum of four different contributions
\begin{align}
j^\mu_{\rm MEC}= j^\mu_{\pi}+j^\mu_{\rm sea}+ j^\mu_{\rm pole}+ j^\mu_{\Delta}\, ,
\end{align}
whose corresponding Feynman diagrams are depicted in Fig.~\ref{mec:diag}. We note that these currents are symmetric under the exchange $1\leftrightarrow 2$.

Introducing the pion momenta $k_1=p-h$ and $k_2=p^\prime - h^\prime$, the 
pion-in-flight current operator corresponding to diagram (a) of Fig.~\ref{mec:diag}
is written as
\begin{align}
j^\mu_\pi & =(I_V)_\pm J^\mu_\pi \ ,\nonumber\\
J^\mu_\pi &=(J^\mu_\pi)_V+(J^\mu_\pi)_A\ ,\nonumber\\
(J^\mu_\pi)_V &= \frac{f^2_{\pi NN}}{m_\pi^2}F_1^V(q) F_{\pi NN}(k_1)F_{\pi NN}(k_2) \nonumber\\
&\times \Pi(k_1)_{(1)}\Pi(k_2)_{(2)}(k_1^\mu-k_2^\mu)\, .
\label{pif:curr}
\end{align}
while its axial component vanishes, $(J^\mu_\pi)_A= 0$. In the above equation, $f^2_{\pi NN}/(4\pi)$=0.08 and the pion propagation and absorption is described by
\begin{align}
\Pi(k)=\frac{\gamma_5\slashed{k}}{k^2-m_\pi^2}\, .
\end{align}
The isospin raising-lowering operator is given by
\begin{align}
(I_V)_\pm=(\tau^{(1)}\times \tau^{(2)})_\pm\ ,
\end{align} 
where $\pm\rightarrow x\pm iy$, where in the electromagnetic case we simply take $\pm\rightarrow z$.

Consistent with the conserved vector current (CVC) hypothesis, the vector part of the pion-in-flight current operator includes the electromagnetic form factor
\begin{align} 
F_1^V(q)=G_E^p(q)-G_E^n(q)\, .
\label{fpi:vec}
\end{align}
The $\pi NN$ coupling is described using a form factor that accounts for the off-shellness of the
pion
\begin{align}
&F_{\pi NN}(k)= \frac{\Lambda_\pi^2-m_\pi^2}{\Lambda_\pi^2-k^2}\label{fpinn}\ ,
\end{align}
where $\Lambda_\pi$=1300 MeV.

The electroweak seagull current operator, given by the sum of diagram (b) of Fig. \ref{mec:diag} and
the one obtained interchanging particles 1 and 2, reads
\begin{align}
j^\mu_{\rm sea}&=(I_V)_\pm J^\mu_{\rm sea} \ ,\nonumber\\
J^\mu_{\rm sea}&=(J^\mu_{\rm sea})_V+(J^\mu_{\rm sea})_A\ ,\nonumber\\
(J^\mu_{\rm sea})_V&= \frac{f^2_{\pi NN}}{m_\pi^2}F_1^V(q) F^2_{\pi NN}(k_1) \Pi(k_1)_{(1)} \big (\gamma_5 \gamma^\mu\big)_{(2)} \nonumber\\
&- (1\leftrightarrow 2)\ ,\nonumber\\
(J^\mu_{\rm sea})_A&=\frac{f^2_{\pi NN}}{m_\pi^2}\frac{1}{g_A}F_\rho(k_2) F^2_{\pi NN}(k_1) \Pi(k_1)_{(1)} (\gamma^\mu\big)_{(2)} \nonumber\\
& - (1\leftrightarrow 2)\, .
\label{sea:curr}
\end{align}
The form factor $F_\rho(k)$, included to account for the $\rho$ meson dominance of the $\pi NN$ coupling,
is given by 
\cite{Hernandez:2007qq}
\begin{align}
F_\rho(k)=\frac{1}{k^2-m^2_\rho}, \ \ \  m_\rho=775.8\ {\rm MeV}
\end{align}

The expression for the pion-pole current operator, represented by diagram (c) of Fig. \ref{mec:diag}, is 
\begin{align}
j^\mu_{\rm pole}&=(I_V)_\pm J^\mu_{\rm pole} \ ,\\
J^\mu_{\rm pole}&=(J^\mu_{\rm pole})_V+(J^\mu_{\rm pole})_A\ , \\
(J^\mu_{\rm pole})_V&=0 ,\\
(J^\mu_{\rm pole})_A&= \frac{f^2_{\pi NN}}{m_\pi^2}\frac{1}{g_A}F_\rho(k_1)F^2_{\pi NN}(k_2) \Pi(k_2)_{(2)} \nonumber\\
&\times  \Big(\frac{q^\mu\slashed{q}}{q^2-m_\pi^2}\Big)_{(1)}- (1\leftrightarrow 2)\ .
\label{pole:curr}
\end{align}

Diagrams (d), as well as the corresponding two in which particles 1 and 2 are interchanged, are associated with
 two-body current terms involving a $\Delta$-resonance in the intermediate state.
The expression of this operator is largely model dependent, owing to the purely transverse nature of this current, \textit{ i.e.}
the form of the vector part  is not subject to current-conservation constraints. We adopted the parametrization of Ref.~\cite{Hernandez:2007qq}
\begin{align}
j^\mu_{\Delta}&=\frac{3}{2}\frac{f_{\pi NN} f^\ast}{m^2_\pi} \bigg\{ \Pi(k_2)_{(2)}
\Big[ \Big( -\frac{2}{3}\tau^{(2)}+\frac{I_V}{3}\Big)_{\pm} \nonumber\\
&\times F_{\pi NN}(k_2) F_{\pi N \Delta} (k_2) (J^\mu_a)_{(1)}-\Big(\frac{2}{3}\tau^{(2)}+\frac{I_V}{3}\Big)_{\pm} \nonumber\\
&\times F_{\pi NN}(k_2) F_{\pi N \Delta} (k_2) (J^\mu_b)_{(1)}\Big]+(1\leftrightarrow 2)\bigg\}
\label{delta:curr}
\end{align}
where $f^\ast$=2.14 and 
\begin{equation}
F_{\pi N \Delta}(k)=\frac{\Lambda^2_{\pi N\Delta}}{\Lambda^2_{\pi N\Delta}-k^2}\ ,
\end{equation}
with $\Lambda_{\pi N\Delta}=1150$ MeV.
The $N\rightarrow \Delta$ transition vertices entering the left and right (d) diagrams, corresponding to $J^\mu_a$ and $J^\mu_b$, respectively
are expressed as
\begin{align}
J^\mu_a&=(J^\mu_a)_V+(J^\mu_a)_A\ ,\nonumber\\
(J^\mu_a)_V&=\frac{C_3^V}{M}\Big[k_2^\alpha G_{\alpha\beta}(h+q)\Big(g^{\beta\mu}\slashed{q}-q^\beta\gamma^\mu\Big)\Big]\gamma_5\ ,\nonumber\\
(J^\mu_a)_A&=C_5^A \Big[k_2^\alpha G_{\alpha\beta}(h+q) g^{\beta\mu}\Big]
\end{align}
and
\begin{align}
J^\mu_b&=(J^\mu_b)_V+(J^\mu_b)_A\ ,\nonumber\\
(J^\mu_b)_V&=\frac{C_3^V}{M}\gamma_5\Big[\Big(g^{\alpha\mu}\slashed{q}-q^\alpha\gamma^\mu\Big)G_{\alpha\beta}(p-q)k_2^\beta\Big],\nonumber\\
(J^\mu_b)_A&=C_5^A \Big[ g^{\alpha\mu} G_{\alpha\beta}(p-q) k_2^\beta\Big]\, .
\end{align}
Since the above $\Delta$ current is applied in the resonance region, the standard Rarita-Schwinger propagator
\begin{align}
G^{\alpha\beta}(p_\Delta)=\frac{P^{\alpha\beta}(p_\Delta)}{p^2_\Delta-M_\Delta^2}
\end{align}
has to be modified to account for the possible $\Delta$  decay into a physical $\pi N$ state. To this aim, following Refs.~\cite{Dekker:1994yc,DePace:2003spn}, 
we replace the real resonance mass $M_\Delta$=1232 MeV by $M_\Delta - i \Gamma(p_\Delta)/2$. The energy-dependent 
decay width $\Gamma(p_\Delta)/2$ effectively accounts for the allowed phase space for the pion produced in the physical decay process. It is
given by
\begin{equation}
\Gamma(p_\Delta)=\frac{(4 f_{\pi N \Delta})^2}{12\pi m_\pi^2} \frac{|\mathbf{k}|^3}{\sqrt{s}} (m_N + E_k) R(\mathbf{r}^2)
\end{equation}
where $(4 f_{\pi N \Delta})^2/(4\pi)=0.38$, $s=p_\Delta^2$ is the invariant mass, $\mathbf{k}$ is the decay three-momentum
in the $\pi N$ center of mass frame, such that
\begin{equation}
\mathbf{k}^2=\frac{1}{4s}[s-(m_N+m_\pi)^2][s-(m_N-m_\pi)^2]\,
\end{equation} 
and $E_k=\sqrt{m_N^2 + \mathbf{k}^2}$ is the associated energy. The additional factor
\begin{equation}
R(\mathbf{r}^2)=\left(\frac{\Lambda_R^2}{\Lambda_R^2-\mathbf{r}^2}\right)
\end{equation}
depending on the $\pi N$ three-momentum $\mathbf{r}$, with $\mathbf{r}^2=(E_k - \sqrt{m_\pi^2 + \mathbf{k}^2})^2-4\mathbf{k}^2$ and $\Lambda_R^2=0.95\, m_N^2$,
is needed to better reproduce the experimental phase-shift $\delta_{33}$~\cite{Dekker:1994yc}.
The medium effects on the $\Delta$ propagator are accounted for by modifying the decay width as
\begin{align}
    \Gamma_\Delta(p_\Delta)\to \Gamma_\Delta(p_\Delta)-2\rm{Im}[U_\Delta(p_\Delta,\rho=\rho_0)],
\end{align}
where $U_\Delta$ is a density dependent potential obtained from a Bruckner-Hartree-Fock calculation using 
a coupled-channel $NN\oplus N\Delta\oplus \pi NN$ model~\cite{Lee:1983xu,Lee:1984us,Lee:1985jq,Lee:1987hd} and we fixed the density at the nuclear saturation value $\rho_0$ =0.16 fm$^3$. A detailed analysis of medium effects in the MEC contribution for electron-nucleus scattering can be found in Ref.~\cite{Rocco:2019gfb}. 
The spin 3/2 projection operator reads
\begin{align}
P^{\alpha\beta}(p_\Delta)&=(\slashed{p}_\Delta+M_\Delta)\Big[ g^\alpha\beta -\frac{1}{3}\gamma^\alpha\gamma^\beta  -\frac{2}{3}\frac{p_\Delta^\alpha p_\Delta^\beta}{M_\Delta^2}\nonumber\\
&+\frac{1}{3}\frac{p_\Delta^\alpha \gamma^\beta - p_\Delta^\beta \gamma^\alpha}{M_\Delta}\Big]\ .
\end{align}

The vector and axial form factors adopted in this work are those of Ref.~\cite{Hernandez:2007qq}
\begin{align}
C_3^V&=\frac{2.13}{(1-q^2/M_V^2)^2}\ \frac{1}{1-q^2/(4 M_V^2)}\ ,\label{cv3:del}\\
C_5^A&=\frac{1.2}{(1-q^2/M^2_{A\Delta})^2}\ \frac{1}{1-q^2/(3M^2_{A\Delta})}\ ,\label{c5a:del}
\end{align}
where $M_V=0.84$ GeV and $M_{A\Delta}=1.05$ GeV. In this work we only keep the leading vector and axial $N\rightarrow\Delta$ form factors, though including the sub-leading terms in the current operator does not pose any conceptual difficulties. 

\section{Interference Contribution}\label{sec:interf}

A fully quantum mechanical treatment of the scattering process requires accounting for the interference between one- and two-body current contributions, potentially leading to both one and two-nucleon knockouts. Within the factorization scheme, calculations based on realistic spectral functions have shown that the contribution to the response tensor from interference leading to two-nucleon knockout is relatively small~\cite{Rocco:2015cil}. Here, we focus on the interference between one- and two-body currents resulting in single-nucleon knockout (one-particle, one-hole) final states. The corresponding response tensor is given by
\begin{equation}\label{eq:intf_R}
    R^{\mu\nu}_{\rm{12b}} = \sum_{f} \langle 0 | j_{2b}^{\mu\dagger}|f\rangle\langle f|j_{1b}^{\nu}|0\rangle \delta(\omega + E_{0} - E_{f})\, .
\end{equation}
Let us initially focus on non-interacting, homogeneous, isospin-symmetric nuclear matter --- a Fermi gas of protons and neutrons. In this case, the two-body current yields a one-particle, one-hole final state, represented by $|ph^{-1}\rangle$,  via the interaction of the struck nucleon with a second spectator nucleon. 

We start by explicitly writing the initial and final nuclear states as properly normalized Slater determinants
\begin{align} 
|ph^{-1}\rangle &=\mathcal{A}[\phi_{\tilde{n}_{1}}\phi_{\tilde{n}_{2}} \dots \phi_{\tilde{n}_{A}}]\nonumber\\
|0\rangle &=\mathcal{A}[\phi_{n_{1}}\phi_{n_{2}} \dots \phi_{n_{A}}]\, .
\end{align} 
In nuclear matter, to respect translation invariance, the single-particle orbitals are plane waves
\begin{equation}
\langle \mathbf{r}_i | \phi_{n}\rangle \equiv \phi_{n}(\mathbf{r}_i)= \frac{e^{i \mathbf{k} \cdot \mathbf{r}_i}}{ \sqrt{V}} |\eta_{n}\rangle \nonumber\, , 
\end{equation}
where the spinor $\eta_{n}$ represents the spin and the isospin of state i. The indexes $\tilde{n}_i$ and ${n}_i$ denote the single particle states of the ground-state and  the one-particle, one-hole final state, respectively. Note that, $\tilde{n}_i= n_i$ except when $n_i = h$, as in that case $\tilde{n}_i = p$.

Exploiting the orthogonality of single-particle orbitals, and the fact that MEC are symmetric under the exchange $1 \leftrightarrow 2$, the many-body matrix element of the reduces to 
\begin{align}
& \langle ph^{-1} |j^{\mu}_{2b}|0 \rangle = \sum_{n}\int d^3r_{1}d^3r_{2} \phi^{*}_{p}(\mathbf{r}_1)\phi^{*}_{n}(\mathbf{r}_2) \nonumber \\
& \times j_{12}(\mathbf{r}_1, \mathbf{r}_2) \mathcal{A}[\phi_{h}(\mathbf{r}_1)\phi_{n}(\mathbf{r}_2)]\, .
\end{align}

As observed in Refs.~\cite{Amaro:1998ta, Amaro:2001xz}, the direct term vanishes for owing to both spin-isospin traces being zero in unpolarized isospin-symmetric nuclear matter, and the presence of a pion carrying zero momentum, thus
\begin{align}
& \langle ph^{-1}|j^\mu_{2b}|0 \rangle = - \sum_{n}\int d^3r_{1}d^3r_{2} \phi^{*}_{p}(\mathbf{r}_1)\phi^{*}_{n}(\mathbf{r}_2) \nonumber \\
& \times j_{12}^\mu(\mathbf{r}_1, \mathbf{r}_2) \phi_{n}(\mathbf{r}_1)\phi_{h}(\mathbf{r}_2)\,.
\label{eq:two_ME_coordinate}
\end{align}
The coordinate space expressions of the two-body current operators are connected to the ones in momentum space provided in Section~\ref{sec:currents} by the following Fourier transformation
\begin{align}
    j^\mu_{12}(\mathbf{r}_1,\mathbf{r}) =& \int \frac{d^{3}k_1}{(2\pi)^{3}}\frac{d^{3}k_2} {(2\pi)^{3}}e^{i\mathbf{k}_1 \cdot \mathbf{r}_1}e^{i \mathbf{k}_2 \cdot \mathbf{r}_2} \nonumber \\
    & (2\pi)^{3}\delta(\mathbf{k}_1 + \mathbf{k}_2 -\mathbf{q})j^\mu_{12}(\mathbf{k}_1,\mathbf{k}_2)\,.
\end{align}
We can now insert the latter expression into Eq.~\eqref{eq:two_ME_coordinate} and integrate over $\mathbf{r}_1$ and $\mathbf{r}_2$. Recalling that
\begin{align}
& \sum_n = V \int \frac{d^3k}{(2\pi)^3} \theta({k_F}-{\bf k})\nonumber\\
& \int d^3 r e^{i \mathbf{k} \cdot \mathbf{r}} = (2\pi)^3 \delta(\mathbf{k})\, ,
\end{align}
and integrating over $\mathbf{k}_1$ and $\mathbf{k}_2$, we obtain 
\begin{align}\label{eq:two_ME_momemtum}
 \langle ph^{-1} |j^{\mu}_{2b}|0 \rangle & =  - \frac{1}{V}\sum_{\eta_k}  \int d^3 k \delta^3({\bf q} - {\bf h} + {\bf p})\nonumber\\
& \times \theta({k_F}-{\bf k})\langle\eta_{p}\eta_{k}|j^\mu_{12}(\mathbf{k}_1, \mathbf{k}_2)|\eta_{h}\eta_{p}\rangle\, .
\end{align}

Following the same steps outlined above, the nuclear-matter matrix element of the one-body current operator can be readily expressed as
\begin{align}
\langle ph^{-1}|j^{\mu}_{1b}|0\rangle &=\frac{1}{V}(2\pi)^3\delta^3({\bf q} - {\bf h} + {\bf p})\langle\eta_{p}|j_{1}|\eta_{h}\rangle\, .
\end{align}

Substituting the expressions just obtained for the one- and two-body current matrix elements into Eq.~\eqref{eq:intf_R}, the interference contribution to the response tensor for the nuclear matter reads
\begin{align}
R^{\mu\nu}_{12b} =& - V \sum_{\eta_p,\eta_h,\eta_k}\int \frac{d^3h}{(2\pi)^3}  \frac{d^3 k}{(2\pi)^3} \Big[\delta({\bf q} - {\bf h} + {\bf p})   \nonumber  \\
& \times \theta({\bf p}-k_F)\theta({k_F}-{\bf h})  \theta({k_F}-{\bf k})\langle\eta_{p}|j_{1}|\eta_{h}\rangle   \nonumber \\
&\times \langle\eta_{p}\eta_{k}|j_{12}|\eta_{h}\eta_{k}\rangle \delta(\omega - e(\mathbf{p}) + e(\mathbf{h}))\Big]\, .
\end{align}

To make contact with finite nuclei, we replace
\begin{equation}
\theta(k_F-{\bf k})  \rightarrow \tilde{n}_{\rm MF}({\bf k})
\end{equation}
and
\begin{align}
& \theta (k_F - {\bf h})\delta (\omega + e({\bf h})- e({\bf p})) \rightarrow \nonumber\\
& \qquad\qquad \int dE {\tilde P}_{\rm MF}({\bf h},E) \delta(\tilde{\omega}+ e({\bf h})- e({\bf p}))\,.
\end{align}
To keep the normalization consistent with the infinite matter case, we defined $\tilde{n}_{\rm MF}({\bf h}) = k_F^3/(6\pi^2) n_{\rm MF}({\bf h})$ and $\tilde{P}_{\rm MF}({\bf h},E) = k_F^3/(6\pi^2) P_{\rm MF}({\bf h},E)$. In order to select final states in the reaction with only one-nucleon emission, we consider just the mean-field component of the hole spectral function. The corresponding momentum distribution is obtained as
\begin{equation}
    n_{\rm MF}(\mathbf{k}) = \int dE P_{\rm MF}(\mathbf{k},E)\, .
\end{equation}
Note that, the normalization of the latter is less than $1$, thereby quenching the interference between one- and two-body currents. The final expression for latter that we adopt in our numerical calculation reads 
\begin{align}
R^{\mu\nu}_{12b}=&-V \sum_{\eta_p,\eta_h,\eta_k}\int \frac{d^3h}{(2\pi)^3}\frac{d^3k}{(2\pi)^3} dE \Big[ {\tilde P}_{\rm MF}({\bf h},E)  \nonumber\\
&\times {\tilde n}_{\rm MF}({\bf k}) \delta({\bf q} - {\bf h} + {\bf p})   \theta({\bf p}-k_F) \langle\eta_{p}|j_{1}|\eta_{h}\rangle \nonumber \\ & \times \langle\eta_{p}\eta_{k}|j_{12}|\eta_{h}\eta_{k}\rangle\delta(\tilde{\omega} - e(\mathbf{p}) + e(\mathbf{h}))\Big]\, .
\end{align}

\begin{figure}[b]
    \includegraphics[width=0.5\textwidth]{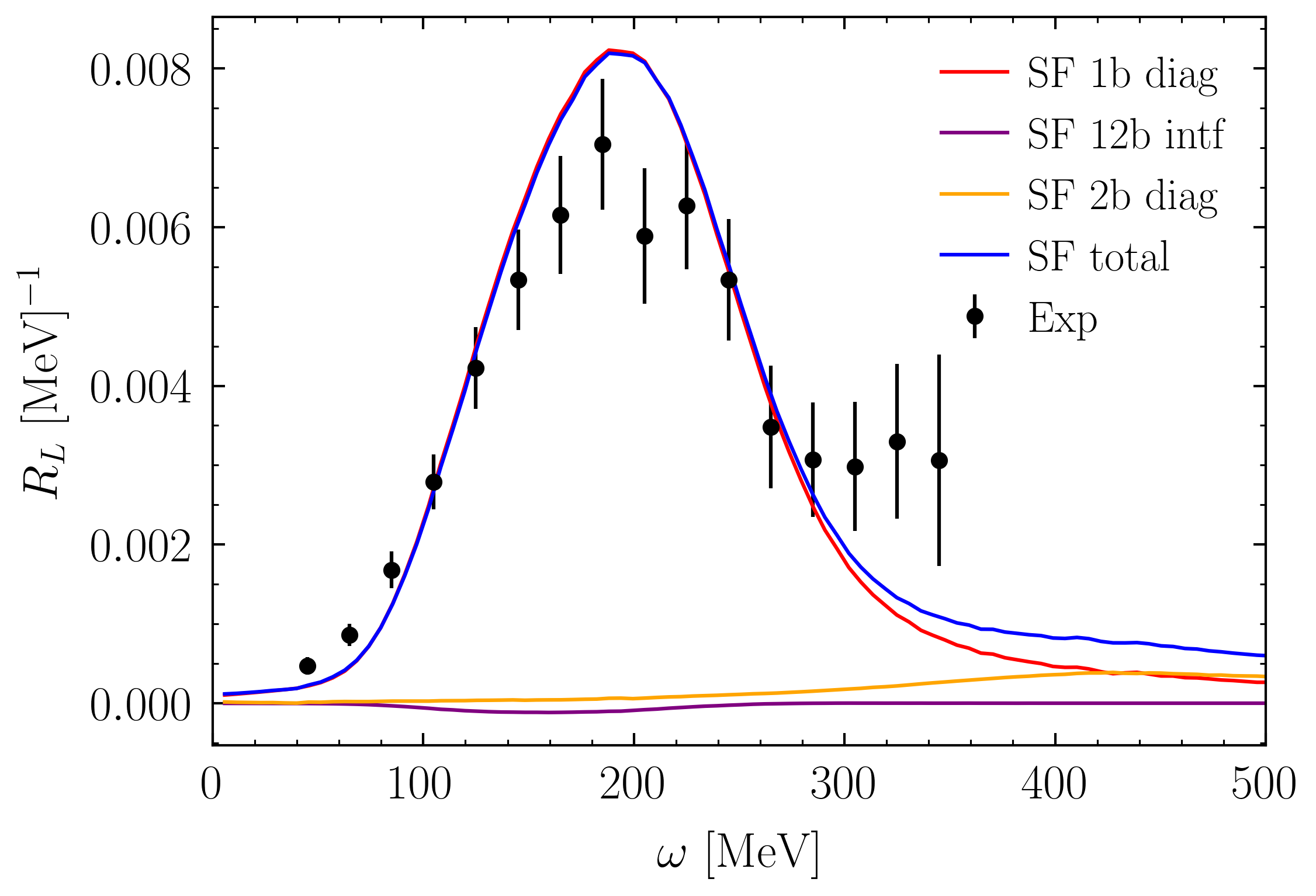} 
    \includegraphics[width=0.5\textwidth]{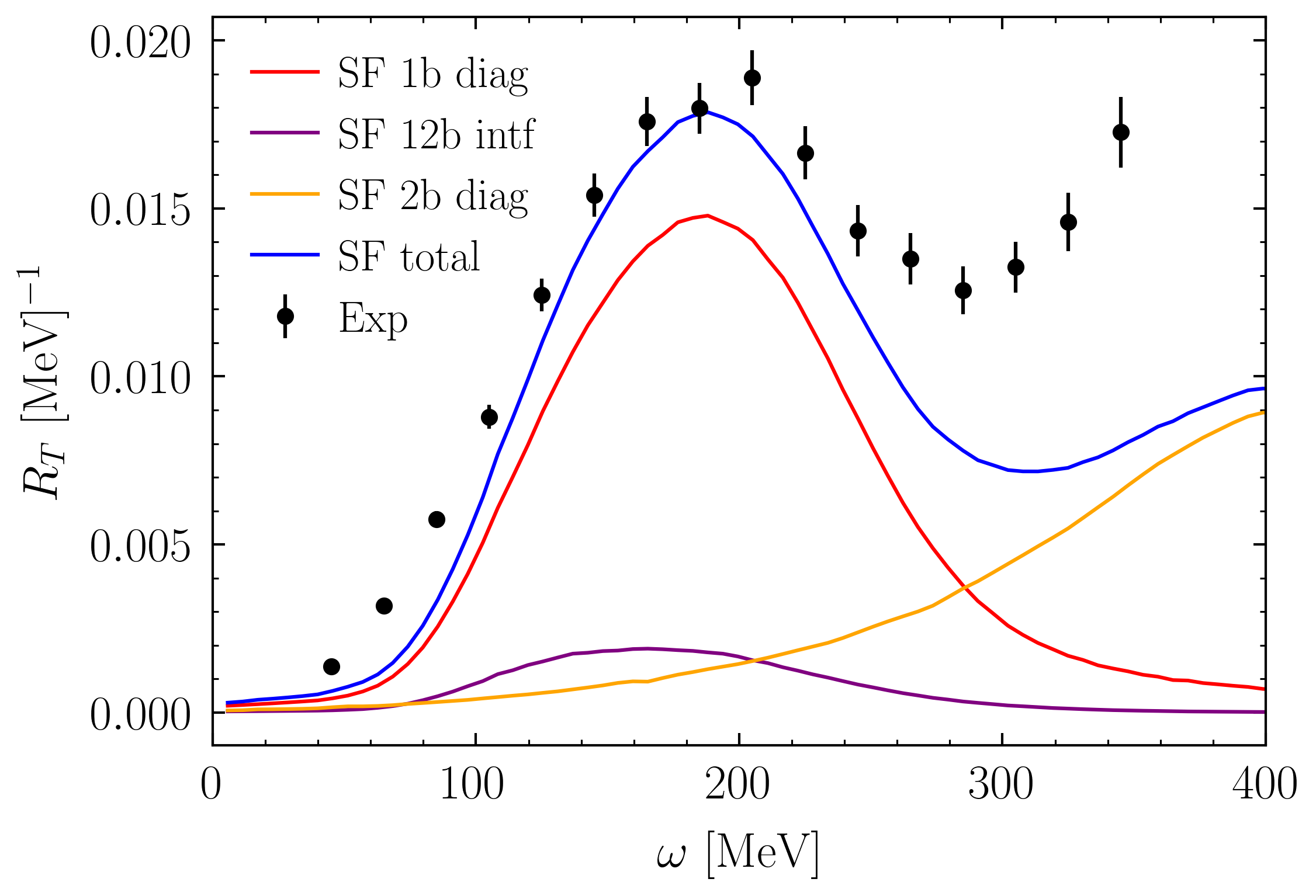} 
    \caption{Longitudinal (upper panel) and transverse (lower panel) response functions from e-$^{12}$C scattering at $|\mathbf{q}| = 570$ MeV. Contributions are separated into pure one-body (red), pure two-body (orange), interference between one and two-body (purple), and total (blue).}
    \label{fig:RLT_responses}
\end{figure}

\section{Results}\label{sec:results}
We begin our analysis by comparing against inclusive electron scattering data, for which the Rosenbluth technique can be applied to separate the longitudinal and transverse contributions. The upper and lower panels of Figure~\ref{fig:RLT_responses} display theoretical calculations of the longitudinal and transverse response functions of $^{12}\rm{C}$, respectively, at a momentum transfer of $|\mathbf{q}| = 570$ MeV compared against the world data analysis conducted by Jourdan~\cite{Jourdan:1996np}. Within the extended factorization scheme, final state interaction effects are included by convoluting the impulse-approximation results with a folding function that both shifts and redistributes strength from the peak to the tails~\cite{Benhar:2006wy}. Contributions from pure one- and two-body currents, as well as their interference, are separately shown to better identify their relative importance.

Consistent with the GFMC calculations in Ref.~\cite{Lovato:2016gkq}, two-body currents in the longitudinal channel appear to have a negligible effect. The pure two-body component is nearly zero in the QE peak, while it brings about a modest enhancement of the response in the dip region. On the other hand, the interference contribution is small in magnitude and negative, resulting in a minor depletion of the total response.
The scenario is markedly different in the transverse channel, where two-body effects play a crucial role. This is not entirely surprising, given that the dominant contribution is the Delta current, which is fully transverse. Both the pure two-body and interference components contribute about the same additional strength to the response beneath the QE peak, constituting almost 20\% of the peak height. This enhancement significantly improves the agreement between theoretical calculations and experimental data.
As highlighted in several previous works --- see for instance Refs.~\cite{amaro2003delta, Rocco:2015cil} --- the pure two-body term becomes dominant in the dip region. It is important to note that the missing strength compared to experimental data has to be ascribed to resonance-production mechanisms~\cite{Rocco:2018mwt}, which have not been accounted for in the present work.

\begin{figure}[h]
    \includegraphics[width=0.5\textwidth]{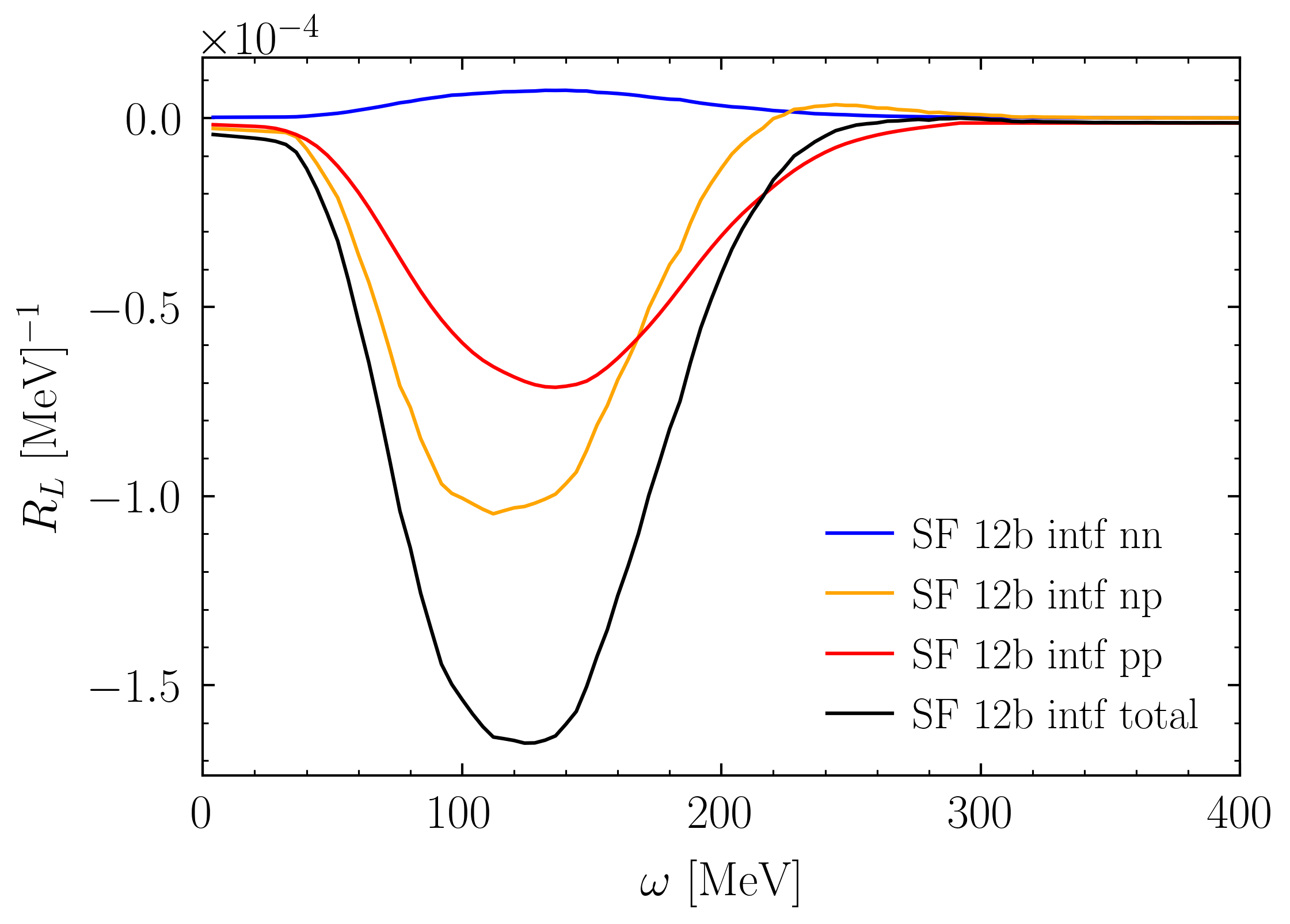} 
    \includegraphics[width=0.5\textwidth]{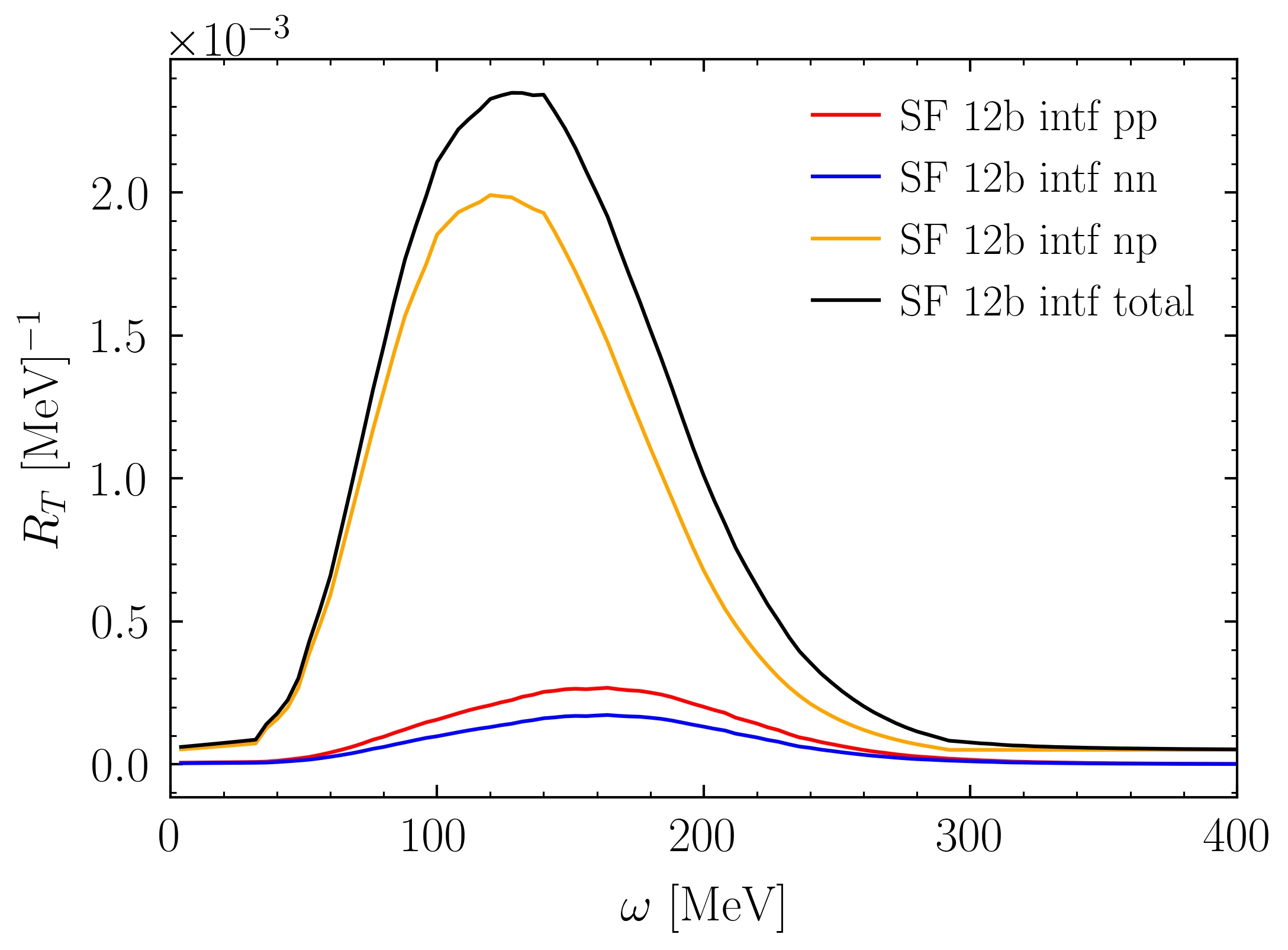} 
    \caption{One- and two-body interference contribution to the longitudinal (upper panel) and transverse (lower panel) response functions from e-$^{12}$C scattering at $|\mathbf{q}| = 500$ MeV separated into isospin pairs: pp (red), np (orange), nn (blue), and total (black).}
    \label{fig:RLT_responses_iso}
\end{figure}

Thanks to our improved numerical implementation of the two-body currents, we are now able to separately compute the contribution of each isospin pair to the nuclear response. Figures~\ref{fig:RLT_responses_iso} show the electromagnetic responses for $|\mathbf{q}| = 500$ MeV, categorized into $pp$, $pn$, and $nn$ pairs. In the transverse response, the contribution of $pn$ pairs is about 10 times as large as that of $pp$ and $nn$ pairs, consistent with the isospin structure observed in SRCs in $(e,e'NN)$ experiments on heavy nuclei~\cite{CLAS:2020rue,CLAS:2018xvc}. The dependence of the pure two-body and interference responses on the isospin of the initial and final state, especially for two-nucleon knockout, is relevant for the MicroBooNE and SBND experiments, as these will constitute a significant fraction of total events~\cite{MicroBooNE:2023cmw}. These experiments aim to measure the outgoing hadronic system with unprecedented accuracy, making semi-exclusive predictions necessary for a correct interpretation of the data. In this context, we note that MicroBooNE has already measured an exclusive sample of $pp$ final state and found large disagreements between model predictions~\cite{MicroBooNE:2022emb}.

\begin{figure*}[htb]
    \includegraphics[width=\textwidth]{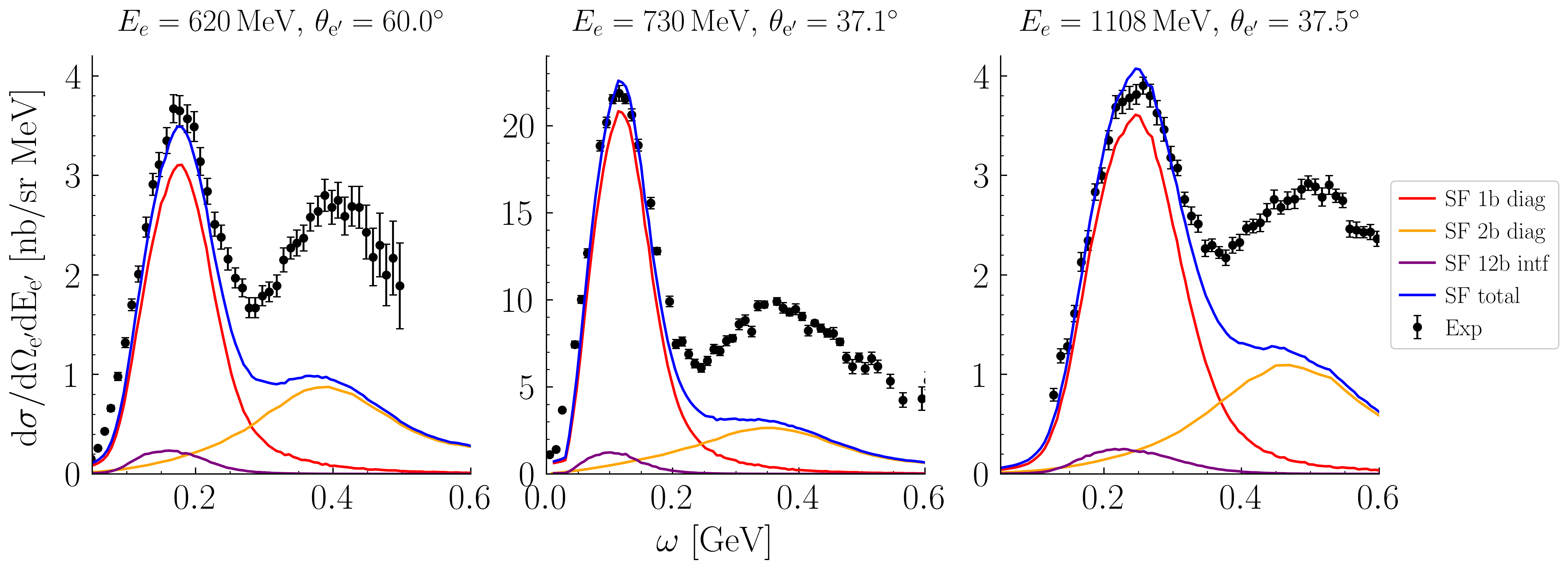} 
    \caption{Inclusive electron cross sections on Carbon at several beam energies and scattering angles. Contributions are separated into pure one-body (red), pure two-body (orange), interference between one and two-body (purple), and total (blue).}
    \label{fig:inclusive_eC}
\end{figure*}

Next, we examine inclusive electron-$^{12}$C cross sections at various beam energies and scattering angles, as depicted in Fig.\ref{fig:inclusive_eC}. The inclusion of one- and two-body interference (indicated by the purple line in this figure) improves the agreement with experimental data by increasing the strength in the QE region by approximately 10\%. Interference effects in the inclusive cross sections appear to be relatively modest, especially for smaller values of $\theta_{e^\prime}$. This behavior can be attributed to the fact that to obtain the inclusive cross section, the longitudinal and transverse responses must be multiplied by their respective leptonic response factors, which depend on the scattering angle. Consequently, two-body currents in general and interference effects, in particular, are more pronounced at backward angles, where the transverse response contributes a significant fraction of the strength, as illustrated in Fig.\ref{fig:inclusive_eC}. Again, the missing strength in the resonance-production region is due to resonance-production processes, which are not the main focus of this work. 

\begin{figure*}[htb]
    \includegraphics[width=\textwidth]{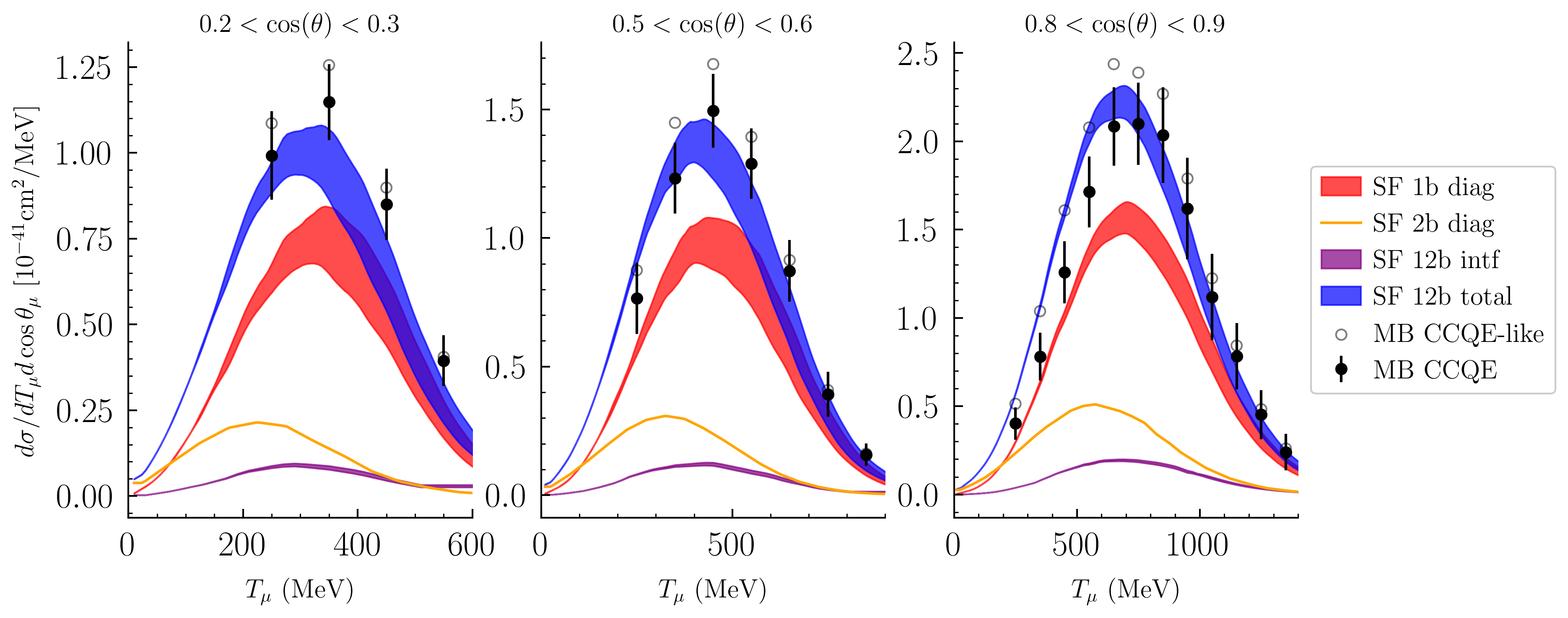} 
    \caption{Flux-averaged $\nu_{\mu}$ differential cross sections on $^{12}$C for MiniBooNE. Three bins of $\cos\theta_{\mu}$ are shown with the one-body contributions in red, pure two-body contributions in orange, one- and two-body interference in purple, and total in blue. The width of the error band interpolates between the dipole axial form factor with $M_{A} = 1$ GeV, and the LQCD form factor of Ref.~\cite{RQCD:2019jai}. The open circles are the cross section to which the background reported in Ref.~\cite{MiniBooNE:2010bsu} is added}
    \label{fig:miniboone}
\end{figure*}

Moving to the electroweak sector, in Fig.\ref{fig:miniboone}, we display the flux-folded charged-current quasielastic (CCQE) $\nu_{\mu}-^{12}$C cross section from MiniBooNE in selected kinematic bins\cite{MiniBooNE:2010bsu}. We present experimental data for both the CCQE-like and the CCQE measurements, with the latter obtained using a model-dependent background subtraction to account for pions created at the interaction vertex and subsequently reabsorbed in the nuclear medium. The difference between the blue (total contribution) and red (one-body currents only) lines highlights the sizable additional strength brought about by two-body currents. As the cross sections are convoluted with the broad MiniBooNE flux, the contribution from interference effects is approximately the same in each bin of $\cos\theta_{\mu}$, constituting about 10\% of the cross section.

The theoretical calculations for the one-body, interference, and total cross sections displayed in Figure~\ref{fig:miniboone} are accompanied by colored bands, whose widths reflect the dependence of the inclusive cross section on different parameterizations of the axial form factor. Specifically, we employ both the dipole parameterization with $M_{A} = 1$ GeV~\cite{Bodek:2007ym}, as extracted from neutrino-deuteron scattering and pion electroproduction experiments, and the LQCD form factor computed in Ref.\cite{RQCD:2019jai}. As extensively discussed, the dipole parameterization is unphysical, and several recent LQCD calculations and measurements of $\bar{\nu}_{\mu}-p$ scattering cross sections tend to favor a form factor with a shallower fall-off for $Q^{2} > 0.3,\rm{GeV}^2$\cite{Meyer:2016oeg,RQCD:2019jai,Park:2021ypf,Djukanovic:2022wru,MINERvA:2023avz,Simons:2022ltq}. 

Theoretical calculations using the LQCD form factor as input consistently yield larger pure one-body cross sections, corresponding to the upper edge of the band. This behavior is particularly evident at backward muon angles, where the $Q^{2}$ dependence of the two axial form factor parameterizations differs more significantly. In contrast, the interference contribution shows a milder dependence on the choice of the axial form factor, increasing by up to 10\% in the most backward angle bin. It is crucial to emphasize that these bands should not be interpreted as the uncertainty in the axial form factor propagated forward to the uncertainty of the cross section but should be seen as a way to interpolate between these two form factor parameterizations.

While the choice of the LQCD form factor seems to significantly improve the agreement with data, the model dependent background subtraction method adopted by the MiniBooNE collaboration as well as the lack of a prediction including events with absorbed pions make quantitative comparisons difficult. We note that in our factorization scheme the enhancement from the LQCD form factor matches the enhancement seen in Green's Function Monte Carlo (GFMC) calculations of flux folded cross section using the same LQCD form factor~\cite{Simons:2022ltq}. As these are two completely different many body methods, only linked by the same underlying nuclear Hamiltonian, the sensitivity to the choice in axial form factor seems robust.

\section{Conclusions}
\label{sec:conclusion}
Providing accurate theoretical predictions, accompanied by reliable uncertainty quantification, for neutrino-nucleus scattering cross-sections in the energy regime relevant to the neutrino-oscillation problem is highly non-trivial. The primary challenges lie in combining a microscopic, quantum-mechanical description of real-time nuclear dynamics with relativistic kinematics and currents. In this regard, the extended factorization scheme, based on realistic spectral functions obtained from Quantum Monte Carlo calculations, offers an ideal framework for addressing this problem. This scheme can accommodate fully relativistic kinematics while maintaining an accurate description of the initial state and the hadronic final state.

To extract neutrino-oscillation parameters with high accuracy and capitalize on the next-generation experiments, it is essential to account for as many quantum-mechanical effects as possible. One such effect is the interference between one- and two-body relativistic current operators, resulting in a final state with one nucleon in the continuum. In this work, we have incorporated this process into the extended factorization scheme in a consistent manner with pure one- and two-body current transitions, ensuring the avoidance of potential double counting.

We first assessed the impact of interference between one- and two-body currents on the electromagnetic longitudinal and transverse response functions. Consistent with GFMC and STA calculations~\cite{Lovato:2016gkq, Pastore:2019urn, Andreoli:2021cxo}, we observed a substantial enhancement of the transverse response in the QE region, while the longitudinal response remained largely unchanged. By analyzing the individual contributions of $pp$, $nn$, and $np$ pairs, we determined that the $np$ pairs predominantly drive the transverse enhancement. This observation is relevant for both electron-scattering and neutrino-oscillation experiments, as it provides insights into the isospin of the emitted nucleon and, consequently, that of the spectator system. Interference effects remain sizable in inclusive electron and neutrino scattering of $^{12}$C, contributing to over 10\% of the cross section, depending on the kinematics. Overall, the inclusion of interference effects in the extended factorization scheme improves the agreement with experimental data for both electron and flux-folded neutrino-nucleus scattering.

Quantifying the uncertainties introduced by approximating the solution of the quantum many-body problem, such as assuming the factorization of the final hadronic state, will become increasingly relevant as statistical and experimental systemic errors diminish in short- and long-baseline experiments. A promising approach to assess the intrinsic errors in the factorization scheme involves comparisons with virtually exact --- albeit non-relativistic --- GFMC calculations. By incorporating interference effects in the extended factorization scheme and addressing relativistic effects in the GFMC through the selection of an appropriate reference frame and the two-fragment model~\cite{Efros:2009qp,Rocco:2018tes,Nikolakopoulos:2023zse}, we can perform more accurate comparisons between these two methods.

Interference effects also play a crucial role in analyzing the axial-vector form factor. Currently, the majority of models rely on the dipole form factor with an axial mass of approximately $1$ GeV, derived from fits to bubble chamber data and pion electro-production~\cite{Bodek:2007ym}. However, this parameterization is recognized to underestimate uncertainty, being both unphysical and inconsistent with QCD sum rules~\cite{Meyer:2016oeg}. In contrast, LQCD computations of single nucleon form factors have advanced to the point where the axial-vector form factor has been computed and accompanied by credible uncertainty estimates. Employing the LQCD form factor of Ref.\cite{RQCD:2019jai} in the extended factorization scheme significantly enhances the QE neutrino-nucleus cross section, bringing it closer to experimental data—although the model dependence on modeling pion reabsorption remains to be fully determined\cite{Meyer:2022mix}.

Event generators typically use the axial mass as a fit parameter in their tunes, extracting it from neutrino-nucleus scattering data. This approach assumes that the underlying nuclear quantum many-body method, which takes the axial form factor as input, is complete, meaning it accurately incorporates all possible interaction channels. However, since most event generators do not include interference effects between one and two-body currents --- which contribute additional strength in the QE region --- these extractions may yield incorrect values for the axial mass. It would be timely to investigate how the inclusion of one- and two-body interference effects affects the extracted value of the axial mass from high-precision neutrino scattering data.

\section{Acknowledgements}
This work was supported in part by the U.S. Department of Energy (DOE), Office of Science, Office of Nuclear Physics under contract number DE-AC02-06CH11357 (A.L.), and by SciDAC-NeuCol (A.L., N.R.)  projects. N.S. is also supported by a Fermilab LDRD award. A.L. is also supported by DOE Early Career Research Program awards. Numerical calculations were performed on the parallel computers of the Laboratory Computing Resource Center, Argonne National Laboratory, the computers of the Argonne Leadership Computing Facility via the INCITE grant ``Ab-initio Nuclear Structure and Nuclear Reactions''.

\bibliography{Interference}

\begin{thebibliography}{83}%
\makeatletter
\providecommand \@ifxundefined [1]{%
 \@ifx{#1\undefined}
}%
\providecommand \@ifnum [1]{%
 \ifnum #1\expandafter \@firstoftwo
 \else \expandafter \@secondoftwo
 \fi
}%
\providecommand \@ifx [1]{%
 \ifx #1\expandafter \@firstoftwo
 \else \expandafter \@secondoftwo
 \fi
}%
\providecommand \natexlab [1]{#1}%
\providecommand \enquote  [1]{``#1''}%
\providecommand \bibnamefont  [1]{#1}%
\providecommand \bibfnamefont [1]{#1}%
\providecommand \citenamefont [1]{#1}%
\providecommand \href@noop [0]{\@secondoftwo}%
\providecommand \href [0]{\begingroup \@sanitize@url \@href}%
\providecommand \@href[1]{\@@startlink{#1}\@@href}%
\providecommand \@@href[1]{\endgroup#1\@@endlink}%
\providecommand \@sanitize@url [0]{\catcode `\\12\catcode `\$12\catcode
  `\&12\catcode `\#12\catcode `\^12\catcode `\_12\catcode `\%12\relax}%
\providecommand \@@startlink[1]{}%
\providecommand \@@endlink[0]{}%
\providecommand \url  [0]{\begingroup\@sanitize@url \@url }%
\providecommand \@url [1]{\endgroup\@href {#1}{\urlprefix }}%
\providecommand \urlprefix  [0]{URL }%
\providecommand \Eprint [0]{\href }%
\providecommand \doibase [0]{http://dx.doi.org/}%
\providecommand \selectlanguage [0]{\@gobble}%
\providecommand \bibinfo  [0]{\@secondoftwo}%
\providecommand \bibfield  [0]{\@secondoftwo}%
\providecommand \translation [1]{[#1]}%
\providecommand \BibitemOpen [0]{}%
\providecommand \bibitemStop [0]{}%
\providecommand \bibitemNoStop [0]{.\EOS\space}%
\providecommand \EOS [0]{\spacefactor3000\relax}%
\providecommand \BibitemShut  [1]{\csname bibitem#1\endcsname}%
\let\auto@bib@innerbib\@empty
\bibitem [{\citenamefont {Acciarri}\ \emph {et~al.}(2015)\citenamefont
  {Acciarri} \emph {et~al.}}]{DUNE:2015lol}%
  \BibitemOpen
  \bibfield  {author} {\bibinfo {author} {\bibfnamefont {R.}~\bibnamefont
  {Acciarri}} \emph {et~al.} (\bibinfo {collaboration} {DUNE}),\ }\href@noop {}
  {\enquote {\bibinfo {title} {{Long-Baseline Neutrino Facility (LBNF) and Deep
  Underground Neutrino Experiment (DUNE)}: {Conceptual Design Report, Volume 2:
  The Physics Program for DUNE at LBNF}},}\ } (\bibinfo {year} {2015}),\
  \Eprint {http://arxiv.org/abs/1512.06148} {arXiv:1512.06148
  [physics.ins-det]} \BibitemShut {NoStop}%
\bibitem [{\citenamefont {Abe}\ \emph {et~al.}(2021{\natexlab{a}})\citenamefont
  {Abe} \emph {et~al.}}]{T2K:2021xwb}%
  \BibitemOpen
  \bibfield  {author} {\bibinfo {author} {\bibfnamefont {K.}~\bibnamefont
  {Abe}} \emph {et~al.} (\bibinfo {collaboration} {T2K}),\ }\href {\doibase
  10.1103/PhysRevD.103.112008} {\bibfield  {journal} {\bibinfo  {journal}
  {Phys. Rev. D}\ }\textbf {\bibinfo {volume} {103}},\ \bibinfo {pages}
  {112008} (\bibinfo {year} {2021}{\natexlab{a}})},\ \Eprint
  {http://arxiv.org/abs/2101.03779} {arXiv:2101.03779 [hep-ex]} \BibitemShut
  {NoStop}%
\bibitem [{\citenamefont {Abe}\ \emph {et~al.}(2018)\citenamefont {Abe} \emph
  {et~al.}}]{Hyper-Kamiokande:2018ofw}%
  \BibitemOpen
  \bibfield  {author} {\bibinfo {author} {\bibfnamefont {K.}~\bibnamefont
  {Abe}} \emph {et~al.} (\bibinfo {collaboration} {Hyper-Kamiokande}),\
  }\href@noop {} {\enquote {\bibinfo {title} {{Hyper-Kamiokande Design
  Report}},}\ } (\bibinfo {year} {2018}),\ \Eprint
  {http://arxiv.org/abs/1805.04163} {arXiv:1805.04163 [physics.ins-det]}
  \BibitemShut {NoStop}%
\bibitem [{\citenamefont {Benhar}\ \emph {et~al.}(2017)\citenamefont {Benhar},
  \citenamefont {Huber}, \citenamefont {Mariani},\ and\ \citenamefont
  {Meloni}}]{Benhar:2015wva}%
  \BibitemOpen
  \bibfield  {author} {\bibinfo {author} {\bibfnamefont {O.}~\bibnamefont
  {Benhar}}, \bibinfo {author} {\bibfnamefont {P.}~\bibnamefont {Huber}},
  \bibinfo {author} {\bibfnamefont {C.}~\bibnamefont {Mariani}}, \ and\
  \bibinfo {author} {\bibfnamefont {D.}~\bibnamefont {Meloni}},\ }\href
  {\doibase 10.1016/j.physrep.2017.07.004} {\bibfield  {journal} {\bibinfo
  {journal} {Phys. Rept.}\ }\textbf {\bibinfo {volume} {700}},\ \bibinfo
  {pages} {1} (\bibinfo {year} {2017})},\ \Eprint
  {http://arxiv.org/abs/1501.06448} {arXiv:1501.06448 [nucl-th]} \BibitemShut
  {NoStop}%
\bibitem [{\citenamefont {Katori}\ and\ \citenamefont
  {Martini}(2018)}]{Katori:2016yel}%
  \BibitemOpen
  \bibfield  {author} {\bibinfo {author} {\bibfnamefont {T.}~\bibnamefont
  {Katori}}\ and\ \bibinfo {author} {\bibfnamefont {M.}~\bibnamefont
  {Martini}},\ }\href {\doibase 10.1088/1361-6471/aa8bf7} {\bibfield  {journal}
  {\bibinfo  {journal} {J. Phys. G}\ }\textbf {\bibinfo {volume} {45}},\
  \bibinfo {pages} {013001} (\bibinfo {year} {2018})},\ \Eprint
  {http://arxiv.org/abs/1611.07770} {arXiv:1611.07770 [hep-ph]} \BibitemShut
  {NoStop}%
\bibitem [{\citenamefont {Alvarez-Ruso}\ \emph {et~al.}(2018)\citenamefont
  {Alvarez-Ruso} \emph {et~al.}}]{NuSTEC:2017hzk}%
  \BibitemOpen
  \bibfield  {author} {\bibinfo {author} {\bibfnamefont {L.}~\bibnamefont
  {Alvarez-Ruso}} \emph {et~al.} (\bibinfo {collaboration} {NuSTEC}),\ }\href
  {\doibase 10.1016/j.ppnp.2018.01.006} {\bibfield  {journal} {\bibinfo
  {journal} {Prog. Part. Nucl. Phys.}\ }\textbf {\bibinfo {volume} {100}},\
  \bibinfo {pages} {1} (\bibinfo {year} {2018})},\ \Eprint
  {http://arxiv.org/abs/1706.03621} {arXiv:1706.03621 [hep-ph]} \BibitemShut
  {NoStop}%
\bibitem [{\citenamefont {Ruso}\ \emph {et~al.}(2022)\citenamefont {Ruso} \emph
  {et~al.}}]{Ruso:2022qes}%
  \BibitemOpen
  \bibfield  {author} {\bibinfo {author} {\bibfnamefont {L.~A.}\ \bibnamefont
  {Ruso}} \emph {et~al.},\ }\href@noop {} {\enquote {\bibinfo {title}
  {{Theoretical tools for neutrino scattering: interplay between lattice QCD,
  EFTs, nuclear physics, phenomenology, and neutrino event generators}},}\ }
  (\bibinfo {year} {2022}),\ \bibinfo {note} {contribution to: Snowmass 2021},\
  \Eprint {http://arxiv.org/abs/2203.09030} {arXiv:2203.09030 [hep-ph]}
  \BibitemShut {NoStop}%
\bibitem [{\citenamefont {Ankowski}\ \emph {et~al.}(2015)\citenamefont
  {Ankowski}, \citenamefont {Benhar}, \citenamefont {Coloma}, \citenamefont
  {Huber}, \citenamefont {Jen}, \citenamefont {Mariani}, \citenamefont
  {Meloni},\ and\ \citenamefont {Vagnoni}}]{Ankowski:2015jya}%
  \BibitemOpen
  \bibfield  {author} {\bibinfo {author} {\bibfnamefont {A.~M.}\ \bibnamefont
  {Ankowski}}, \bibinfo {author} {\bibfnamefont {O.}~\bibnamefont {Benhar}},
  \bibinfo {author} {\bibfnamefont {P.}~\bibnamefont {Coloma}}, \bibinfo
  {author} {\bibfnamefont {P.}~\bibnamefont {Huber}}, \bibinfo {author}
  {\bibfnamefont {C.-M.}\ \bibnamefont {Jen}}, \bibinfo {author} {\bibfnamefont
  {C.}~\bibnamefont {Mariani}}, \bibinfo {author} {\bibfnamefont
  {D.}~\bibnamefont {Meloni}}, \ and\ \bibinfo {author} {\bibfnamefont
  {E.}~\bibnamefont {Vagnoni}},\ }\href {\doibase 10.1103/PhysRevD.92.073014}
  {\bibfield  {journal} {\bibinfo  {journal} {Phys. Rev. D}\ }\textbf {\bibinfo
  {volume} {92}},\ \bibinfo {pages} {073014} (\bibinfo {year} {2015})},\
  \Eprint {http://arxiv.org/abs/1507.08560} {arXiv:1507.08560 [hep-ph]}
  \BibitemShut {NoStop}%
\bibitem [{\citenamefont {Carneiro}\ \emph {et~al.}(2020)\citenamefont
  {Carneiro} \emph {et~al.}}]{MINERvA:2019gsf}%
  \BibitemOpen
  \bibfield  {author} {\bibinfo {author} {\bibfnamefont {M.~F.}\ \bibnamefont
  {Carneiro}} \emph {et~al.} (\bibinfo {collaboration} {MINERvA}),\ }\href
  {\doibase 10.1103/PhysRevLett.124.121801} {\bibfield  {journal} {\bibinfo
  {journal} {Phys. Rev. Lett.}\ }\textbf {\bibinfo {volume} {124}},\ \bibinfo
  {pages} {121801} (\bibinfo {year} {2020})},\ \Eprint
  {http://arxiv.org/abs/1912.09890} {arXiv:1912.09890 [hep-ex]} \BibitemShut
  {NoStop}%
\bibitem [{\citenamefont {Abe}\ \emph {et~al.}(2020)\citenamefont {Abe} \emph
  {et~al.}}]{T2K:2020jav}%
  \BibitemOpen
  \bibfield  {author} {\bibinfo {author} {\bibfnamefont {K.}~\bibnamefont
  {Abe}} \emph {et~al.} (\bibinfo {collaboration} {T2K}),\ }\href {\doibase
  10.1103/PhysRevD.101.112004} {\bibfield  {journal} {\bibinfo  {journal}
  {Phys. Rev. D}\ }\textbf {\bibinfo {volume} {101}},\ \bibinfo {pages}
  {112004} (\bibinfo {year} {2020})},\ \Eprint
  {http://arxiv.org/abs/2004.05434} {arXiv:2004.05434 [hep-ex]} \BibitemShut
  {NoStop}%
\bibitem [{\citenamefont {Khachatryan}\ \emph {et~al.}(2021)\citenamefont
  {Khachatryan} \emph {et~al.}}]{CLAS:2021neh}%
  \BibitemOpen
  \bibfield  {author} {\bibinfo {author} {\bibfnamefont {M.}~\bibnamefont
  {Khachatryan}} \emph {et~al.} (\bibinfo {collaboration} {CLAS, e4v}),\ }\href
  {\doibase 10.1038/s41586-021-04046-5} {\bibfield  {journal} {\bibinfo
  {journal} {Nature}\ }\textbf {\bibinfo {volume} {599}},\ \bibinfo {pages}
  {565} (\bibinfo {year} {2021})}\BibitemShut {NoStop}%
\bibitem [{\citenamefont {Cai}\ \emph {et~al.}(2020)\citenamefont {Cai} \emph
  {et~al.}}]{MINERvA:2019ope}%
  \BibitemOpen
  \bibfield  {author} {\bibinfo {author} {\bibfnamefont {T.}~\bibnamefont
  {Cai}} \emph {et~al.} (\bibinfo {collaboration} {MINERvA}),\ }\href {\doibase
  10.1103/PhysRevD.101.092001} {\bibfield  {journal} {\bibinfo  {journal}
  {Phys. Rev. D}\ }\textbf {\bibinfo {volume} {101}},\ \bibinfo {pages}
  {092001} (\bibinfo {year} {2020})},\ \Eprint
  {http://arxiv.org/abs/1910.08658} {arXiv:1910.08658 [hep-ex]} \BibitemShut
  {NoStop}%
\bibitem [{\citenamefont {Ascencio}\ \emph {et~al.}(2022)\citenamefont
  {Ascencio} \emph {et~al.}}]{MINERvA:2021wjs}%
  \BibitemOpen
  \bibfield  {author} {\bibinfo {author} {\bibfnamefont {M.~V.}\ \bibnamefont
  {Ascencio}} \emph {et~al.} (\bibinfo {collaboration} {MINERvA}),\ }\href
  {\doibase 10.1103/PhysRevD.106.032001} {\bibfield  {journal} {\bibinfo
  {journal} {Phys. Rev. D}\ }\textbf {\bibinfo {volume} {106}},\ \bibinfo
  {pages} {032001} (\bibinfo {year} {2022})},\ \Eprint
  {http://arxiv.org/abs/2110.13372} {arXiv:2110.13372 [hep-ex]} \BibitemShut
  {NoStop}%
\bibitem [{\citenamefont {Abe}\ \emph {et~al.}(2021{\natexlab{b}})\citenamefont
  {Abe} \emph {et~al.}}]{T2K:2021naz}%
  \BibitemOpen
  \bibfield  {author} {\bibinfo {author} {\bibfnamefont {K.}~\bibnamefont
  {Abe}} \emph {et~al.} (\bibinfo {collaboration} {T2K}),\ }\href {\doibase
  10.1103/PhysRevD.103.112009} {\bibfield  {journal} {\bibinfo  {journal}
  {Phys. Rev. D}\ }\textbf {\bibinfo {volume} {103}},\ \bibinfo {pages}
  {112009} (\bibinfo {year} {2021}{\natexlab{b}})},\ \Eprint
  {http://arxiv.org/abs/2102.03346} {arXiv:2102.03346 [hep-ex]} \BibitemShut
  {NoStop}%
\bibitem [{\citenamefont {Abratenko}\ \emph
  {et~al.}(2023{\natexlab{a}})\citenamefont {Abratenko} \emph
  {et~al.}}]{MicroBooNE:2023cmw}%
  \BibitemOpen
  \bibfield  {author} {\bibinfo {author} {\bibfnamefont {P.}~\bibnamefont
  {Abratenko}} \emph {et~al.} (\bibinfo {collaboration} {MicroBooNE}),\ }\href
  {\doibase 10.1103/PhysRevD.108.053002} {\bibfield  {journal} {\bibinfo
  {journal} {Phys. Rev. D}\ }\textbf {\bibinfo {volume} {108}},\ \bibinfo
  {pages} {053002} (\bibinfo {year} {2023}{\natexlab{a}})},\ \Eprint
  {http://arxiv.org/abs/2301.03700} {arXiv:2301.03700 [hep-ex]} \BibitemShut
  {NoStop}%
\bibitem [{\citenamefont {Abratenko}\ \emph {et~al.}(2022)\citenamefont
  {Abratenko} \emph {et~al.}}]{MicroBooNE:2022emb}%
  \BibitemOpen
  \bibfield  {author} {\bibinfo {author} {\bibfnamefont {P.}~\bibnamefont
  {Abratenko}} \emph {et~al.} (\bibinfo {collaboration} {MicroBooNE}),\
  }\href@noop {} {\  (\bibinfo {year} {2022})},\ \Eprint
  {http://arxiv.org/abs/2211.03734} {arXiv:2211.03734 [hep-ex]} \BibitemShut
  {NoStop}%
\bibitem [{\citenamefont {Ruterbories}\ \emph {et~al.}(2022)\citenamefont
  {Ruterbories} \emph {et~al.}}]{MINERvA:2022mnw}%
  \BibitemOpen
  \bibfield  {author} {\bibinfo {author} {\bibfnamefont {D.}~\bibnamefont
  {Ruterbories}} \emph {et~al.} (\bibinfo {collaboration} {MINERvA}),\ }\href
  {\doibase 10.1103/PhysRevLett.129.021803} {\bibfield  {journal} {\bibinfo
  {journal} {Phys. Rev. Lett.}\ }\textbf {\bibinfo {volume} {129}},\ \bibinfo
  {pages} {021803} (\bibinfo {year} {2022})},\ \Eprint
  {http://arxiv.org/abs/2203.08022} {arXiv:2203.08022 [hep-ex]} \BibitemShut
  {NoStop}%
\bibitem [{\citenamefont {Abratenko}\ \emph
  {et~al.}(2023{\natexlab{b}})\citenamefont {Abratenko} \emph
  {et~al.}}]{MicroBooNE:2023tzj}%
  \BibitemOpen
  \bibfield  {author} {\bibinfo {author} {\bibfnamefont {P.}~\bibnamefont
  {Abratenko}} \emph {et~al.} (\bibinfo {collaboration} {MicroBooNE}),\ }\href
  {\doibase 10.1103/PhysRevLett.131.101802} {\bibfield  {journal} {\bibinfo
  {journal} {Phys. Rev. Lett.}\ }\textbf {\bibinfo {volume} {131}},\ \bibinfo
  {pages} {101802} (\bibinfo {year} {2023}{\natexlab{b}})},\ \Eprint
  {http://arxiv.org/abs/2301.03706} {arXiv:2301.03706 [hep-ex]} \BibitemShut
  {NoStop}%
\bibitem [{\citenamefont {Kleykamp}\ \emph {et~al.}(2023)\citenamefont
  {Kleykamp} \emph {et~al.}}]{MINERvA:2023kuz}%
  \BibitemOpen
  \bibfield  {author} {\bibinfo {author} {\bibfnamefont {J.}~\bibnamefont
  {Kleykamp}} \emph {et~al.} (\bibinfo {collaboration} {MINERvA}),\ }\href
  {\doibase 10.1103/PhysRevLett.130.161801} {\bibfield  {journal} {\bibinfo
  {journal} {Phys. Rev. Lett.}\ }\textbf {\bibinfo {volume} {130}},\ \bibinfo
  {pages} {161801} (\bibinfo {year} {2023})},\ \Eprint
  {http://arxiv.org/abs/2301.02272} {arXiv:2301.02272 [hep-ex]} \BibitemShut
  {NoStop}%
\bibitem [{\citenamefont {Abi}\ \emph {et~al.}(2021)\citenamefont {Abi} \emph
  {et~al.}}]{DUNE:2020fgq}%
  \BibitemOpen
  \bibfield  {author} {\bibinfo {author} {\bibfnamefont {B.}~\bibnamefont
  {Abi}} \emph {et~al.} (\bibinfo {collaboration} {DUNE}),\ }\href {\doibase
  10.1140/epjc/s10052-021-09007-w} {\bibfield  {journal} {\bibinfo  {journal}
  {Eur. Phys. J. C}\ }\textbf {\bibinfo {volume} {81}},\ \bibinfo {pages} {322}
  (\bibinfo {year} {2021})},\ \Eprint {http://arxiv.org/abs/2008.12769}
  {arXiv:2008.12769 [hep-ex]} \BibitemShut {NoStop}%
\bibitem [{\citenamefont {Carlson}\ \emph {et~al.}(2015)\citenamefont
  {Carlson}, \citenamefont {Gandolfi}, \citenamefont {Pederiva}, \citenamefont
  {Pieper}, \citenamefont {Schiavilla}, \citenamefont {Schmidt},\ and\
  \citenamefont {Wiringa}}]{Carlson:2014vla}%
  \BibitemOpen
  \bibfield  {author} {\bibinfo {author} {\bibfnamefont {J.}~\bibnamefont
  {Carlson}}, \bibinfo {author} {\bibfnamefont {S.}~\bibnamefont {Gandolfi}},
  \bibinfo {author} {\bibfnamefont {F.}~\bibnamefont {Pederiva}}, \bibinfo
  {author} {\bibfnamefont {S.~C.}\ \bibnamefont {Pieper}}, \bibinfo {author}
  {\bibfnamefont {R.}~\bibnamefont {Schiavilla}}, \bibinfo {author}
  {\bibfnamefont {K.~E.}\ \bibnamefont {Schmidt}}, \ and\ \bibinfo {author}
  {\bibfnamefont {R.~B.}\ \bibnamefont {Wiringa}},\ }\href {\doibase
  10.1103/RevModPhys.87.1067} {\bibfield  {journal} {\bibinfo  {journal} {Rev.
  Mod. Phys.}\ }\textbf {\bibinfo {volume} {87}},\ \bibinfo {pages} {1067}
  (\bibinfo {year} {2015})},\ \Eprint {http://arxiv.org/abs/1412.3081}
  {arXiv:1412.3081 [nucl-th]} \BibitemShut {NoStop}%
\bibitem [{\citenamefont {Gandolfi}\ \emph {et~al.}(2020)\citenamefont
  {Gandolfi}, \citenamefont {Lonardoni}, \citenamefont {Lovato},\ and\
  \citenamefont {Piarulli}}]{Gandolfi:2020pbj}%
  \BibitemOpen
  \bibfield  {author} {\bibinfo {author} {\bibfnamefont {S.}~\bibnamefont
  {Gandolfi}}, \bibinfo {author} {\bibfnamefont {D.}~\bibnamefont {Lonardoni}},
  \bibinfo {author} {\bibfnamefont {A.}~\bibnamefont {Lovato}}, \ and\ \bibinfo
  {author} {\bibfnamefont {M.}~\bibnamefont {Piarulli}},\ }\href {\doibase
  10.3389/fphy.2020.00117} {\bibfield  {journal} {\bibinfo  {journal} {Front.
  in Phys.}\ }\textbf {\bibinfo {volume} {8}},\ \bibinfo {pages} {117}
  (\bibinfo {year} {2020})},\ \Eprint {http://arxiv.org/abs/2001.01374}
  {arXiv:2001.01374 [nucl-th]} \BibitemShut {NoStop}%
\bibitem [{\citenamefont {Lovato}\ \emph {et~al.}(2020)\citenamefont {Lovato},
  \citenamefont {Carlson}, \citenamefont {Gandolfi}, \citenamefont {Rocco},\
  and\ \citenamefont {Schiavilla}}]{Lovato:2020kba}%
  \BibitemOpen
  \bibfield  {author} {\bibinfo {author} {\bibfnamefont {A.}~\bibnamefont
  {Lovato}}, \bibinfo {author} {\bibfnamefont {J.}~\bibnamefont {Carlson}},
  \bibinfo {author} {\bibfnamefont {S.}~\bibnamefont {Gandolfi}}, \bibinfo
  {author} {\bibfnamefont {N.}~\bibnamefont {Rocco}}, \ and\ \bibinfo {author}
  {\bibfnamefont {R.}~\bibnamefont {Schiavilla}},\ }\href {\doibase
  10.1103/PhysRevX.10.031068} {\bibfield  {journal} {\bibinfo  {journal} {Phys.
  Rev. X}\ }\textbf {\bibinfo {volume} {10}},\ \bibinfo {pages} {031068}
  (\bibinfo {year} {2020})},\ \Eprint {http://arxiv.org/abs/2003.07710}
  {arXiv:2003.07710 [nucl-th]} \BibitemShut {NoStop}%
\bibitem [{\citenamefont {Lovato}\ \emph {et~al.}(2018)\citenamefont {Lovato},
  \citenamefont {Gandolfi}, \citenamefont {Carlson}, \citenamefont {Lusk},
  \citenamefont {Pieper},\ and\ \citenamefont {Schiavilla}}]{Lovato:2017cux}%
  \BibitemOpen
  \bibfield  {author} {\bibinfo {author} {\bibfnamefont {A.}~\bibnamefont
  {Lovato}}, \bibinfo {author} {\bibfnamefont {S.}~\bibnamefont {Gandolfi}},
  \bibinfo {author} {\bibfnamefont {J.}~\bibnamefont {Carlson}}, \bibinfo
  {author} {\bibfnamefont {E.}~\bibnamefont {Lusk}}, \bibinfo {author}
  {\bibfnamefont {S.~C.}\ \bibnamefont {Pieper}}, \ and\ \bibinfo {author}
  {\bibfnamefont {R.}~\bibnamefont {Schiavilla}},\ }\href {\doibase
  10.1103/PhysRevC.97.022502} {\bibfield  {journal} {\bibinfo  {journal} {Phys.
  Rev. C}\ }\textbf {\bibinfo {volume} {97}},\ \bibinfo {pages} {022502}
  (\bibinfo {year} {2018})},\ \Eprint {http://arxiv.org/abs/1711.02047}
  {arXiv:1711.02047 [nucl-th]} \BibitemShut {NoStop}%
\bibitem [{\citenamefont {Lovato}\ \emph {et~al.}(2016)\citenamefont {Lovato},
  \citenamefont {Gandolfi}, \citenamefont {Carlson}, \citenamefont {Pieper},\
  and\ \citenamefont {Schiavilla}}]{Lovato:2016gkq}%
  \BibitemOpen
  \bibfield  {author} {\bibinfo {author} {\bibfnamefont {A.}~\bibnamefont
  {Lovato}}, \bibinfo {author} {\bibfnamefont {S.}~\bibnamefont {Gandolfi}},
  \bibinfo {author} {\bibfnamefont {J.}~\bibnamefont {Carlson}}, \bibinfo
  {author} {\bibfnamefont {S.~C.}\ \bibnamefont {Pieper}}, \ and\ \bibinfo
  {author} {\bibfnamefont {R.}~\bibnamefont {Schiavilla}},\ }\href {\doibase
  10.1103/PhysRevLett.117.082501} {\bibfield  {journal} {\bibinfo  {journal}
  {Phys. Rev. Lett.}\ }\textbf {\bibinfo {volume} {117}},\ \bibinfo {pages}
  {082501} (\bibinfo {year} {2016})},\ \Eprint
  {http://arxiv.org/abs/1605.00248} {arXiv:1605.00248 [nucl-th]} \BibitemShut
  {NoStop}%
\bibitem [{\citenamefont {Sobczyk}\ \emph {et~al.}(2023)\citenamefont
  {Sobczyk}, \citenamefont {Acharya}, \citenamefont {Bacca},\ and\
  \citenamefont {Hagen}}]{Sobczyk:2023sxh}%
  \BibitemOpen
  \bibfield  {author} {\bibinfo {author} {\bibfnamefont {J.~E.}\ \bibnamefont
  {Sobczyk}}, \bibinfo {author} {\bibfnamefont {B.}~\bibnamefont {Acharya}},
  \bibinfo {author} {\bibfnamefont {S.}~\bibnamefont {Bacca}}, \ and\ \bibinfo
  {author} {\bibfnamefont {G.}~\bibnamefont {Hagen}},\ }\href@noop {} {\
  (\bibinfo {year} {2023})},\ \Eprint {http://arxiv.org/abs/2310.03109}
  {arXiv:2310.03109 [nucl-th]} \BibitemShut {NoStop}%
\bibitem [{\citenamefont {Ekstr\"om}\ \emph {et~al.}(2015)\citenamefont
  {Ekstr\"om}, \citenamefont {Jansen}, \citenamefont {Wendt}, \citenamefont
  {Hagen}, \citenamefont {Papenbrock}, \citenamefont {Carlsson}, \citenamefont
  {Forss\'en}, \citenamefont {Hjorth-Jensen}, \citenamefont {Navr\'atil},\ and\
  \citenamefont {Nazarewicz}}]{Ekstrom:2015rta}%
  \BibitemOpen
  \bibfield  {author} {\bibinfo {author} {\bibfnamefont {A.}~\bibnamefont
  {Ekstr\"om}}, \bibinfo {author} {\bibfnamefont {G.~R.}\ \bibnamefont
  {Jansen}}, \bibinfo {author} {\bibfnamefont {K.~A.}\ \bibnamefont {Wendt}},
  \bibinfo {author} {\bibfnamefont {G.}~\bibnamefont {Hagen}}, \bibinfo
  {author} {\bibfnamefont {T.}~\bibnamefont {Papenbrock}}, \bibinfo {author}
  {\bibfnamefont {B.~D.}\ \bibnamefont {Carlsson}}, \bibinfo {author}
  {\bibfnamefont {C.}~\bibnamefont {Forss\'en}}, \bibinfo {author}
  {\bibfnamefont {M.}~\bibnamefont {Hjorth-Jensen}}, \bibinfo {author}
  {\bibfnamefont {P.}~\bibnamefont {Navr\'atil}}, \ and\ \bibinfo {author}
  {\bibfnamefont {W.}~\bibnamefont {Nazarewicz}},\ }\href {\doibase
  10.1103/PhysRevC.91.051301} {\bibfield  {journal} {\bibinfo  {journal} {Phys.
  Rev. C}\ }\textbf {\bibinfo {volume} {91}},\ \bibinfo {pages} {051301}
  (\bibinfo {year} {2015})},\ \Eprint {http://arxiv.org/abs/1502.04682}
  {arXiv:1502.04682 [nucl-th]} \BibitemShut {NoStop}%
\bibitem [{\citenamefont {Jiang}\ \emph {et~al.}(2020)\citenamefont {Jiang},
  \citenamefont {Ekstr\"om}, \citenamefont {Forss\'en}, \citenamefont {Hagen},
  \citenamefont {Jansen},\ and\ \citenamefont {Papenbrock}}]{Jiang:2020the}%
  \BibitemOpen
  \bibfield  {author} {\bibinfo {author} {\bibfnamefont {W.~G.}\ \bibnamefont
  {Jiang}}, \bibinfo {author} {\bibfnamefont {A.}~\bibnamefont {Ekstr\"om}},
  \bibinfo {author} {\bibfnamefont {C.}~\bibnamefont {Forss\'en}}, \bibinfo
  {author} {\bibfnamefont {G.}~\bibnamefont {Hagen}}, \bibinfo {author}
  {\bibfnamefont {G.~R.}\ \bibnamefont {Jansen}}, \ and\ \bibinfo {author}
  {\bibfnamefont {T.}~\bibnamefont {Papenbrock}},\ }\href {\doibase
  10.1103/PhysRevC.102.054301} {\bibfield  {journal} {\bibinfo  {journal}
  {Phys. Rev. C}\ }\textbf {\bibinfo {volume} {102}},\ \bibinfo {pages}
  {054301} (\bibinfo {year} {2020})},\ \Eprint
  {http://arxiv.org/abs/2006.16774} {arXiv:2006.16774 [nucl-th]} \BibitemShut
  {NoStop}%
\bibitem [{\citenamefont {Rocco}\ \emph {et~al.}(2018)\citenamefont {Rocco},
  \citenamefont {Leidemann}, \citenamefont {Lovato},\ and\ \citenamefont
  {Orlandini}}]{Rocco:2018tes}%
  \BibitemOpen
  \bibfield  {author} {\bibinfo {author} {\bibfnamefont {N.}~\bibnamefont
  {Rocco}}, \bibinfo {author} {\bibfnamefont {W.}~\bibnamefont {Leidemann}},
  \bibinfo {author} {\bibfnamefont {A.}~\bibnamefont {Lovato}}, \ and\ \bibinfo
  {author} {\bibfnamefont {G.}~\bibnamefont {Orlandini}},\ }\href {\doibase
  10.1103/PhysRevC.97.055501} {\bibfield  {journal} {\bibinfo  {journal} {Phys.
  Rev. C}\ }\textbf {\bibinfo {volume} {97}},\ \bibinfo {pages} {055501}
  (\bibinfo {year} {2018})},\ \Eprint {http://arxiv.org/abs/1801.07111}
  {arXiv:1801.07111 [nucl-th]} \BibitemShut {NoStop}%
\bibitem [{\citenamefont {Nikolakopoulos}\ \emph {et~al.}(2023)\citenamefont
  {Nikolakopoulos}, \citenamefont {Lovato},\ and\ \citenamefont
  {Rocco}}]{Nikolakopoulos:2023zse}%
  \BibitemOpen
  \bibfield  {author} {\bibinfo {author} {\bibfnamefont {A.}~\bibnamefont
  {Nikolakopoulos}}, \bibinfo {author} {\bibfnamefont {A.}~\bibnamefont
  {Lovato}}, \ and\ \bibinfo {author} {\bibfnamefont {N.}~\bibnamefont
  {Rocco}},\ }\href@noop {} {\  (\bibinfo {year} {2023})},\ \Eprint
  {http://arxiv.org/abs/2304.11772} {arXiv:2304.11772 [nucl-th]} \BibitemShut
  {NoStop}%
\bibitem [{\citenamefont {Abratenko}\ \emph {et~al.}(2021)\citenamefont
  {Abratenko} \emph {et~al.}}]{MicroBooNE:2021ddy}%
  \BibitemOpen
  \bibfield  {author} {\bibinfo {author} {\bibfnamefont {P.}~\bibnamefont
  {Abratenko}} \emph {et~al.} (\bibinfo {collaboration} {MicroBooNE}),\ }\href
  {\doibase 10.1007/JHEP12(2021)153} {\bibfield  {journal} {\bibinfo  {journal}
  {JHEP}\ }\textbf {\bibinfo {volume} {12}},\ \bibinfo {pages} {153} (\bibinfo
  {year} {2021})},\ \Eprint {http://arxiv.org/abs/2109.02460} {arXiv:2109.02460
  [physics.ins-det]} \BibitemShut {NoStop}%
\bibitem [{\citenamefont {Castiglioni}\ \emph {et~al.}(2020)\citenamefont
  {Castiglioni}, \citenamefont {Foreman}, \citenamefont {Lepetic},
  \citenamefont {Littlejohn}, \citenamefont {Malaker},\ and\ \citenamefont
  {Mastbaum}}]{Castiglioni:2020tsu}%
  \BibitemOpen
  \bibfield  {author} {\bibinfo {author} {\bibfnamefont {W.}~\bibnamefont
  {Castiglioni}}, \bibinfo {author} {\bibfnamefont {W.}~\bibnamefont
  {Foreman}}, \bibinfo {author} {\bibfnamefont {I.}~\bibnamefont {Lepetic}},
  \bibinfo {author} {\bibfnamefont {B.~R.}\ \bibnamefont {Littlejohn}},
  \bibinfo {author} {\bibfnamefont {M.}~\bibnamefont {Malaker}}, \ and\
  \bibinfo {author} {\bibfnamefont {A.}~\bibnamefont {Mastbaum}},\ }\href
  {\doibase 10.1103/PhysRevD.102.092010} {\bibfield  {journal} {\bibinfo
  {journal} {Phys. Rev. D}\ }\textbf {\bibinfo {volume} {102}},\ \bibinfo
  {pages} {092010} (\bibinfo {year} {2020})},\ \Eprint
  {http://arxiv.org/abs/2006.14675} {arXiv:2006.14675 [physics.ins-det]}
  \BibitemShut {NoStop}%
\bibitem [{\citenamefont {Benhar}\ and\ \citenamefont
  {Pandharipande}(1993)}]{Benhar:1993ja}%
  \BibitemOpen
  \bibfield  {author} {\bibinfo {author} {\bibfnamefont {O.}~\bibnamefont
  {Benhar}}\ and\ \bibinfo {author} {\bibfnamefont {V.~R.}\ \bibnamefont
  {Pandharipande}},\ }\href {\doibase 10.1103/PhysRevC.47.2218} {\bibfield
  {journal} {\bibinfo  {journal} {Phys. Rev. C}\ }\textbf {\bibinfo {volume}
  {47}},\ \bibinfo {pages} {2218} (\bibinfo {year} {1993})}\BibitemShut
  {NoStop}%
\bibitem [{\citenamefont {Benhar}\ \emph {et~al.}(2008)\citenamefont {Benhar},
  \citenamefont {day},\ and\ \citenamefont {Sick}}]{Benhar:2006wy}%
  \BibitemOpen
  \bibfield  {author} {\bibinfo {author} {\bibfnamefont {O.}~\bibnamefont
  {Benhar}}, \bibinfo {author} {\bibfnamefont {D.}~\bibnamefont {day}}, \ and\
  \bibinfo {author} {\bibfnamefont {I.}~\bibnamefont {Sick}},\ }\href {\doibase
  10.1103/RevModPhys.80.189} {\bibfield  {journal} {\bibinfo  {journal} {Rev.
  Mod. Phys.}\ }\textbf {\bibinfo {volume} {80}},\ \bibinfo {pages} {189}
  (\bibinfo {year} {2008})},\ \Eprint {http://arxiv.org/abs/nucl-ex/0603029}
  {arXiv:nucl-ex/0603029} \BibitemShut {NoStop}%
\bibitem [{\citenamefont {Rocco}(2020)}]{Rocco:2020jlx}%
  \BibitemOpen
  \bibfield  {author} {\bibinfo {author} {\bibfnamefont {N.}~\bibnamefont
  {Rocco}},\ }\href {\doibase 10.3389/fphy.2020.00116} {\bibfield  {journal}
  {\bibinfo  {journal} {Front. in Phys.}\ }\textbf {\bibinfo {volume} {8}},\
  \bibinfo {pages} {116} (\bibinfo {year} {2020})}\BibitemShut {NoStop}%
\bibitem [{\citenamefont {Pastore}\ \emph {et~al.}(2020)\citenamefont
  {Pastore}, \citenamefont {Carlson}, \citenamefont {Gandolfi}, \citenamefont
  {Schiavilla},\ and\ \citenamefont {Wiringa}}]{Pastore:2019urn}%
  \BibitemOpen
  \bibfield  {author} {\bibinfo {author} {\bibfnamefont {S.}~\bibnamefont
  {Pastore}}, \bibinfo {author} {\bibfnamefont {J.}~\bibnamefont {Carlson}},
  \bibinfo {author} {\bibfnamefont {S.}~\bibnamefont {Gandolfi}}, \bibinfo
  {author} {\bibfnamefont {R.}~\bibnamefont {Schiavilla}}, \ and\ \bibinfo
  {author} {\bibfnamefont {R.~B.}\ \bibnamefont {Wiringa}},\ }\href {\doibase
  10.1103/PhysRevC.101.044612} {\bibfield  {journal} {\bibinfo  {journal}
  {Phys. Rev. C}\ }\textbf {\bibinfo {volume} {101}},\ \bibinfo {pages}
  {044612} (\bibinfo {year} {2020})},\ \Eprint
  {http://arxiv.org/abs/1909.06400} {arXiv:1909.06400 [nucl-th]} \BibitemShut
  {NoStop}%
\bibitem [{\citenamefont {Korover}\ \emph {et~al.}(2023)\citenamefont {Korover}
  \emph {et~al.}}]{CLAS:2022odn}%
  \BibitemOpen
  \bibfield  {author} {\bibinfo {author} {\bibfnamefont {I.}~\bibnamefont
  {Korover}} \emph {et~al.} (\bibinfo {collaboration} {CLAS}),\ }\href
  {\doibase 10.1103/PhysRevC.107.L061301} {\bibfield  {journal} {\bibinfo
  {journal} {Phys. Rev. C}\ }\textbf {\bibinfo {volume} {107}},\ \bibinfo
  {pages} {L061301} (\bibinfo {year} {2023})},\ \Eprint
  {http://arxiv.org/abs/2209.01492} {arXiv:2209.01492 [nucl-ex]} \BibitemShut
  {NoStop}%
\bibitem [{\citenamefont {Andreoli}\ \emph {et~al.}(2022)\citenamefont
  {Andreoli}, \citenamefont {Carlson}, \citenamefont {Lovato}, \citenamefont
  {Pastore}, \citenamefont {Rocco},\ and\ \citenamefont
  {Wiringa}}]{Andreoli:2021cxo}%
  \BibitemOpen
  \bibfield  {author} {\bibinfo {author} {\bibfnamefont {L.}~\bibnamefont
  {Andreoli}}, \bibinfo {author} {\bibfnamefont {J.}~\bibnamefont {Carlson}},
  \bibinfo {author} {\bibfnamefont {A.}~\bibnamefont {Lovato}}, \bibinfo
  {author} {\bibfnamefont {S.}~\bibnamefont {Pastore}}, \bibinfo {author}
  {\bibfnamefont {N.}~\bibnamefont {Rocco}}, \ and\ \bibinfo {author}
  {\bibfnamefont {R.~B.}\ \bibnamefont {Wiringa}},\ }\href {\doibase
  10.1103/PhysRevC.105.014002} {\bibfield  {journal} {\bibinfo  {journal}
  {Phys. Rev. C}\ }\textbf {\bibinfo {volume} {105}},\ \bibinfo {pages}
  {014002} (\bibinfo {year} {2022})},\ \Eprint
  {http://arxiv.org/abs/2108.10824} {arXiv:2108.10824 [nucl-th]} \BibitemShut
  {NoStop}%
\bibitem [{\citenamefont {Efros}\ \emph {et~al.}(2010)\citenamefont {Efros},
  \citenamefont {Leidemann}, \citenamefont {Orlandini},\ and\ \citenamefont
  {Tomusiak}}]{Efros:2009qp}%
  \BibitemOpen
  \bibfield  {author} {\bibinfo {author} {\bibfnamefont {V.~D.}\ \bibnamefont
  {Efros}}, \bibinfo {author} {\bibfnamefont {W.}~\bibnamefont {Leidemann}},
  \bibinfo {author} {\bibfnamefont {G.}~\bibnamefont {Orlandini}}, \ and\
  \bibinfo {author} {\bibfnamefont {E.~L.}\ \bibnamefont {Tomusiak}},\ }\href
  {\doibase 10.1103/PhysRevC.81.034001} {\bibfield  {journal} {\bibinfo
  {journal} {Phys. Rev. C}\ }\textbf {\bibinfo {volume} {81}},\ \bibinfo
  {pages} {034001} (\bibinfo {year} {2010})},\ \Eprint
  {http://arxiv.org/abs/0910.2406} {arXiv:0910.2406 [nucl-th]} \BibitemShut
  {NoStop}%
\bibitem [{\citenamefont {Benhar}\ \emph {et~al.}(2015)\citenamefont {Benhar},
  \citenamefont {Lovato},\ and\ \citenamefont {Rocco}}]{Benhar:2015ula}%
  \BibitemOpen
  \bibfield  {author} {\bibinfo {author} {\bibfnamefont {O.}~\bibnamefont
  {Benhar}}, \bibinfo {author} {\bibfnamefont {A.}~\bibnamefont {Lovato}}, \
  and\ \bibinfo {author} {\bibfnamefont {N.}~\bibnamefont {Rocco}},\ }\href
  {\doibase 10.1103/PhysRevC.92.024602} {\bibfield  {journal} {\bibinfo
  {journal} {Phys. Rev. C}\ }\textbf {\bibinfo {volume} {92}},\ \bibinfo
  {pages} {024602} (\bibinfo {year} {2015})},\ \Eprint
  {http://arxiv.org/abs/1502.00887} {arXiv:1502.00887 [nucl-th]} \BibitemShut
  {NoStop}%
\bibitem [{\citenamefont {Rocco}\ \emph {et~al.}(2016)\citenamefont {Rocco},
  \citenamefont {Lovato},\ and\ \citenamefont {Benhar}}]{Rocco:2015cil}%
  \BibitemOpen
  \bibfield  {author} {\bibinfo {author} {\bibfnamefont {N.}~\bibnamefont
  {Rocco}}, \bibinfo {author} {\bibfnamefont {A.}~\bibnamefont {Lovato}}, \
  and\ \bibinfo {author} {\bibfnamefont {O.}~\bibnamefont {Benhar}},\ }\href
  {\doibase 10.1103/PhysRevLett.116.192501} {\bibfield  {journal} {\bibinfo
  {journal} {Phys. Rev. Lett.}\ }\textbf {\bibinfo {volume} {116}},\ \bibinfo
  {pages} {192501} (\bibinfo {year} {2016})},\ \Eprint
  {http://arxiv.org/abs/1512.07426} {arXiv:1512.07426 [nucl-th]} \BibitemShut
  {NoStop}%
\bibitem [{\citenamefont {Franco-Munoz}\ \emph {et~al.}(2022)\citenamefont
  {Franco-Munoz}, \citenamefont {Gonz\'alez-Jim\'enez},\ and\ \citenamefont
  {Ud\'\i{}as}}]{Franco-Munoz:2022jcl}%
  \BibitemOpen
  \bibfield  {author} {\bibinfo {author} {\bibfnamefont {T.}~\bibnamefont
  {Franco-Munoz}}, \bibinfo {author} {\bibfnamefont {R.}~\bibnamefont
  {Gonz\'alez-Jim\'enez}}, \ and\ \bibinfo {author} {\bibfnamefont {J.~M.}\
  \bibnamefont {Ud\'\i{}as}},\ }\href@noop {} {\  (\bibinfo {year} {2022})},\
  \Eprint {http://arxiv.org/abs/2203.09996} {arXiv:2203.09996 [nucl-th]}
  \BibitemShut {NoStop}%
\bibitem [{\citenamefont {Franco-Munoz}\ \emph {et~al.}(2023)\citenamefont
  {Franco-Munoz}, \citenamefont {Garc\'\i{}a-Marcos}, \citenamefont
  {Gonz\'alez-Jim\'enez},\ and\ \citenamefont
  {Ud\'\i{}as}}]{Franco-Munoz:2023zoa}%
  \BibitemOpen
  \bibfield  {author} {\bibinfo {author} {\bibfnamefont {T.}~\bibnamefont
  {Franco-Munoz}}, \bibinfo {author} {\bibfnamefont {J.}~\bibnamefont
  {Garc\'\i{}a-Marcos}}, \bibinfo {author} {\bibfnamefont {R.}~\bibnamefont
  {Gonz\'alez-Jim\'enez}}, \ and\ \bibinfo {author} {\bibfnamefont {J.~M.}\
  \bibnamefont {Ud\'\i{}as}},\ }\href@noop {} {\  (\bibinfo {year} {2023})},\
  \Eprint {http://arxiv.org/abs/2306.10823} {arXiv:2306.10823 [nucl-th]}
  \BibitemShut {NoStop}%
\bibitem [{\citenamefont {Fabrocini}(1997)}]{Fabrocini:1996bu}%
  \BibitemOpen
  \bibfield  {author} {\bibinfo {author} {\bibfnamefont {A.}~\bibnamefont
  {Fabrocini}},\ }\href {\doibase 10.1103/PhysRevC.55.338} {\bibfield
  {journal} {\bibinfo  {journal} {Phys. Rev. C}\ }\textbf {\bibinfo {volume}
  {55}},\ \bibinfo {pages} {338} (\bibinfo {year} {1997})},\ \Eprint
  {http://arxiv.org/abs/nucl-th/9609027} {arXiv:nucl-th/9609027} \BibitemShut
  {NoStop}%
\bibitem [{\citenamefont {Umino}\ and\ \citenamefont
  {Udias}(1995)}]{Umino:1995bql}%
  \BibitemOpen
  \bibfield  {author} {\bibinfo {author} {\bibfnamefont {Y.}~\bibnamefont
  {Umino}}\ and\ \bibinfo {author} {\bibfnamefont {J.~M.}\ \bibnamefont
  {Udias}},\ }\href {\doibase 10.1103/PhysRevC.52.3399} {\bibfield  {journal}
  {\bibinfo  {journal} {Phys. Rev. C}\ }\textbf {\bibinfo {volume} {52}},\
  \bibinfo {pages} {3399} (\bibinfo {year} {1995})},\ \Eprint
  {http://arxiv.org/abs/nucl-th/9602003} {arXiv:nucl-th/9602003} \BibitemShut
  {NoStop}%
\bibitem [{\citenamefont {Amaro}\ \emph
  {et~al.}(2002{\natexlab{a}})\citenamefont {Amaro}, \citenamefont {Barbaro},
  \citenamefont {Caballero}, \citenamefont {Donnelly},\ and\ \citenamefont
  {Molinari}}]{Amaro:2002mj}%
  \BibitemOpen
  \bibfield  {author} {\bibinfo {author} {\bibfnamefont {J.~E.}\ \bibnamefont
  {Amaro}}, \bibinfo {author} {\bibfnamefont {M.~B.}\ \bibnamefont {Barbaro}},
  \bibinfo {author} {\bibfnamefont {J.~A.}\ \bibnamefont {Caballero}}, \bibinfo
  {author} {\bibfnamefont {T.~W.}\ \bibnamefont {Donnelly}}, \ and\ \bibinfo
  {author} {\bibfnamefont {A.}~\bibnamefont {Molinari}},\ }\href {\doibase
  10.1016/S0370-1573(02)00195-3} {\bibfield  {journal} {\bibinfo  {journal}
  {Phys. Rept.}\ }\textbf {\bibinfo {volume} {368}},\ \bibinfo {pages} {317}
  (\bibinfo {year} {2002}{\natexlab{a}})},\ \Eprint
  {http://arxiv.org/abs/nucl-th/0204001} {arXiv:nucl-th/0204001} \BibitemShut
  {NoStop}%
\bibitem [{\citenamefont {Rocco}\ \emph
  {et~al.}(2019{\natexlab{a}})\citenamefont {Rocco}, \citenamefont {Nakamura},
  \citenamefont {Lee},\ and\ \citenamefont {Lovato}}]{Rocco:2019gfb}%
  \BibitemOpen
  \bibfield  {author} {\bibinfo {author} {\bibfnamefont {N.}~\bibnamefont
  {Rocco}}, \bibinfo {author} {\bibfnamefont {S.~X.}\ \bibnamefont {Nakamura}},
  \bibinfo {author} {\bibfnamefont {T.~S.~H.}\ \bibnamefont {Lee}}, \ and\
  \bibinfo {author} {\bibfnamefont {A.}~\bibnamefont {Lovato}},\ }\href
  {\doibase 10.1103/PhysRevC.100.045503} {\bibfield  {journal} {\bibinfo
  {journal} {Phys. Rev. C}\ }\textbf {\bibinfo {volume} {100}},\ \bibinfo
  {pages} {045503} (\bibinfo {year} {2019}{\natexlab{a}})},\ \Eprint
  {http://arxiv.org/abs/1907.01093} {arXiv:1907.01093 [nucl-th]} \BibitemShut
  {NoStop}%
\bibitem [{\citenamefont {Abe}\ \emph {et~al.}(2023)\citenamefont {Abe} \emph
  {et~al.}}]{T2K:2023smv}%
  \BibitemOpen
  \bibfield  {author} {\bibinfo {author} {\bibfnamefont {K.}~\bibnamefont
  {Abe}} \emph {et~al.} (\bibinfo {collaboration} {T2K}),\ }\href {\doibase
  10.1140/epjc/s10052-023-11819-x} {\bibfield  {journal} {\bibinfo  {journal}
  {Eur. Phys. J. C}\ }\textbf {\bibinfo {volume} {83}},\ \bibinfo {pages} {782}
  (\bibinfo {year} {2023})},\ \Eprint {http://arxiv.org/abs/2303.03222}
  {arXiv:2303.03222 [hep-ex]} \BibitemShut {NoStop}%
\bibitem [{\citenamefont {Machado}\ \emph {et~al.}(2019)\citenamefont
  {Machado}, \citenamefont {Palamara},\ and\ \citenamefont
  {Schmitz}}]{Machado:2019oxb}%
  \BibitemOpen
  \bibfield  {author} {\bibinfo {author} {\bibfnamefont {P.~A.}\ \bibnamefont
  {Machado}}, \bibinfo {author} {\bibfnamefont {O.}~\bibnamefont {Palamara}}, \
  and\ \bibinfo {author} {\bibfnamefont {D.~W.}\ \bibnamefont {Schmitz}},\
  }\href {\doibase 10.1146/annurev-nucl-101917-020949} {\bibfield  {journal}
  {\bibinfo  {journal} {Ann. Rev. Nucl. Part. Sci.}\ }\textbf {\bibinfo
  {volume} {69}},\ \bibinfo {pages} {363} (\bibinfo {year} {2019})},\ \Eprint
  {http://arxiv.org/abs/1903.04608} {arXiv:1903.04608 [hep-ex]} \BibitemShut
  {NoStop}%
\bibitem [{\citenamefont {Lovato}\ \emph {et~al.}(2023)\citenamefont {Lovato},
  \citenamefont {Nikolakopoulos}, \citenamefont {Rocco},\ and\ \citenamefont
  {Steinberg}}]{Lovato:2023raf}%
  \BibitemOpen
  \bibfield  {author} {\bibinfo {author} {\bibfnamefont {A.}~\bibnamefont
  {Lovato}}, \bibinfo {author} {\bibfnamefont {A.}~\bibnamefont
  {Nikolakopoulos}}, \bibinfo {author} {\bibfnamefont {N.}~\bibnamefont
  {Rocco}}, \ and\ \bibinfo {author} {\bibfnamefont {N.}~\bibnamefont
  {Steinberg}},\ }\href {\doibase 10.3390/universe9080367} {\bibfield
  {journal} {\bibinfo  {journal} {Universe}\ }\textbf {\bibinfo {volume} {9}},\
  \bibinfo {pages} {367} (\bibinfo {year} {2023})},\ \Eprint
  {http://arxiv.org/abs/2308.00736} {arXiv:2308.00736 [nucl-th]} \BibitemShut
  {NoStop}%
\bibitem [{\citenamefont {Park}\ \emph {et~al.}(2022)\citenamefont {Park},
  \citenamefont {Gupta}, \citenamefont {Yoon}, \citenamefont {Mondal},
  \citenamefont {Bhattacharya}, \citenamefont {Jang}, \citenamefont {Jo\'o},\
  and\ \citenamefont {Winter}}]{Park:2021ypf}%
  \BibitemOpen
  \bibfield  {author} {\bibinfo {author} {\bibfnamefont {S.}~\bibnamefont
  {Park}}, \bibinfo {author} {\bibfnamefont {R.}~\bibnamefont {Gupta}},
  \bibinfo {author} {\bibfnamefont {B.}~\bibnamefont {Yoon}}, \bibinfo {author}
  {\bibfnamefont {S.}~\bibnamefont {Mondal}}, \bibinfo {author} {\bibfnamefont
  {T.}~\bibnamefont {Bhattacharya}}, \bibinfo {author} {\bibfnamefont {Y.-C.}\
  \bibnamefont {Jang}}, \bibinfo {author} {\bibfnamefont {B.}~\bibnamefont
  {Jo\'o}}, \ and\ \bibinfo {author} {\bibfnamefont {F.}~\bibnamefont {Winter}}
  (\bibinfo {collaboration} {Nucleon Matrix Elements (NME)}),\ }\href {\doibase
  10.1103/PhysRevD.105.054505} {\bibfield  {journal} {\bibinfo  {journal}
  {Phys. Rev. D}\ }\textbf {\bibinfo {volume} {105}},\ \bibinfo {pages}
  {054505} (\bibinfo {year} {2022})},\ \Eprint
  {http://arxiv.org/abs/2103.05599} {arXiv:2103.05599 [hep-lat]} \BibitemShut
  {NoStop}%
\bibitem [{\citenamefont {Djukanovic}\ \emph {et~al.}(2022)\citenamefont
  {Djukanovic}, \citenamefont {von Hippel}, \citenamefont {Koponen},
  \citenamefont {Meyer}, \citenamefont {Ottnad}, \citenamefont {Schulz},\ and\
  \citenamefont {Wittig}}]{Djukanovic:2022wru}%
  \BibitemOpen
  \bibfield  {author} {\bibinfo {author} {\bibfnamefont {D.}~\bibnamefont
  {Djukanovic}}, \bibinfo {author} {\bibfnamefont {G.}~\bibnamefont {von
  Hippel}}, \bibinfo {author} {\bibfnamefont {J.}~\bibnamefont {Koponen}},
  \bibinfo {author} {\bibfnamefont {H.~B.}\ \bibnamefont {Meyer}}, \bibinfo
  {author} {\bibfnamefont {K.}~\bibnamefont {Ottnad}}, \bibinfo {author}
  {\bibfnamefont {T.}~\bibnamefont {Schulz}}, \ and\ \bibinfo {author}
  {\bibfnamefont {H.}~\bibnamefont {Wittig}},\ }\href {\doibase
  10.1103/PhysRevD.106.074503} {\bibfield  {journal} {\bibinfo  {journal}
  {Phys. Rev. D}\ }\textbf {\bibinfo {volume} {106}},\ \bibinfo {pages}
  {074503} (\bibinfo {year} {2022})},\ \Eprint
  {http://arxiv.org/abs/2207.03440} {arXiv:2207.03440 [hep-lat]} \BibitemShut
  {NoStop}%
\bibitem [{\citenamefont {Bali}\ \emph {et~al.}(2020)\citenamefont {Bali},
  \citenamefont {Barca}, \citenamefont {Collins}, \citenamefont {Gruber},
  \citenamefont {L\"offler}, \citenamefont {Sch\"afer}, \citenamefont
  {S\"oldner}, \citenamefont {Wein}, \citenamefont {Weish\"aupl},\ and\
  \citenamefont {Wurm}}]{RQCD:2019jai}%
  \BibitemOpen
  \bibfield  {author} {\bibinfo {author} {\bibfnamefont {G.~S.}\ \bibnamefont
  {Bali}}, \bibinfo {author} {\bibfnamefont {L.}~\bibnamefont {Barca}},
  \bibinfo {author} {\bibfnamefont {S.}~\bibnamefont {Collins}}, \bibinfo
  {author} {\bibfnamefont {M.}~\bibnamefont {Gruber}}, \bibinfo {author}
  {\bibfnamefont {M.}~\bibnamefont {L\"offler}}, \bibinfo {author}
  {\bibfnamefont {A.}~\bibnamefont {Sch\"afer}}, \bibinfo {author}
  {\bibfnamefont {W.}~\bibnamefont {S\"oldner}}, \bibinfo {author}
  {\bibfnamefont {P.}~\bibnamefont {Wein}}, \bibinfo {author} {\bibfnamefont
  {S.}~\bibnamefont {Weish\"aupl}}, \ and\ \bibinfo {author} {\bibfnamefont
  {T.}~\bibnamefont {Wurm}} (\bibinfo {collaboration} {RQCD}),\ }\href
  {\doibase 10.1007/JHEP05(2020)126} {\bibfield  {journal} {\bibinfo  {journal}
  {JHEP}\ }\textbf {\bibinfo {volume} {05}},\ \bibinfo {pages} {126} (\bibinfo
  {year} {2020})},\ \Eprint {http://arxiv.org/abs/1911.13150} {arXiv:1911.13150
  [hep-lat]} \BibitemShut {NoStop}%
\bibitem [{\citenamefont {Simons}\ \emph {et~al.}(2022)\citenamefont {Simons},
  \citenamefont {Steinberg}, \citenamefont {Lovato}, \citenamefont {Meurice},
  \citenamefont {Rocco},\ and\ \citenamefont {Wagman}}]{Simons:2022ltq}%
  \BibitemOpen
  \bibfield  {author} {\bibinfo {author} {\bibfnamefont {D.}~\bibnamefont
  {Simons}}, \bibinfo {author} {\bibfnamefont {N.}~\bibnamefont {Steinberg}},
  \bibinfo {author} {\bibfnamefont {A.}~\bibnamefont {Lovato}}, \bibinfo
  {author} {\bibfnamefont {Y.}~\bibnamefont {Meurice}}, \bibinfo {author}
  {\bibfnamefont {N.}~\bibnamefont {Rocco}}, \ and\ \bibinfo {author}
  {\bibfnamefont {M.}~\bibnamefont {Wagman}},\ }\href@noop {} {\  (\bibinfo
  {year} {2022})},\ \Eprint {http://arxiv.org/abs/2210.02455} {arXiv:2210.02455
  [hep-ph]} \BibitemShut {NoStop}%
\bibitem [{\citenamefont {Dekker}\ \emph {et~al.}(1994)\citenamefont {Dekker},
  \citenamefont {Brussaard},\ and\ \citenamefont {Tjon}}]{Dekker:1994yc}%
  \BibitemOpen
  \bibfield  {author} {\bibinfo {author} {\bibfnamefont {M.~J.}\ \bibnamefont
  {Dekker}}, \bibinfo {author} {\bibfnamefont {P.~J.}\ \bibnamefont
  {Brussaard}}, \ and\ \bibinfo {author} {\bibfnamefont {J.~A.}\ \bibnamefont
  {Tjon}},\ }\href {\doibase 10.1103/PhysRevC.49.2650} {\bibfield  {journal}
  {\bibinfo  {journal} {Phys. Rev. C}\ }\textbf {\bibinfo {volume} {49}},\
  \bibinfo {pages} {2650} (\bibinfo {year} {1994})}\BibitemShut {NoStop}%
\bibitem [{\citenamefont {De~Pace}\ \emph {et~al.}(2003)\citenamefont
  {De~Pace}, \citenamefont {Nardi}, \citenamefont {Alberico}, \citenamefont
  {Donnelly},\ and\ \citenamefont {Molinari}}]{DePace:2003spn}%
  \BibitemOpen
  \bibfield  {author} {\bibinfo {author} {\bibfnamefont {A.}~\bibnamefont
  {De~Pace}}, \bibinfo {author} {\bibfnamefont {M.}~\bibnamefont {Nardi}},
  \bibinfo {author} {\bibfnamefont {W.~M.}\ \bibnamefont {Alberico}}, \bibinfo
  {author} {\bibfnamefont {T.~W.}\ \bibnamefont {Donnelly}}, \ and\ \bibinfo
  {author} {\bibfnamefont {A.}~\bibnamefont {Molinari}},\ }\href {\doibase
  10.1016/S0375-9474(03)01625-7} {\bibfield  {journal} {\bibinfo  {journal}
  {Nucl. Phys. A}\ }\textbf {\bibinfo {volume} {726}},\ \bibinfo {pages} {303}
  (\bibinfo {year} {2003})},\ \Eprint {http://arxiv.org/abs/nucl-th/0304084}
  {arXiv:nucl-th/0304084} \BibitemShut {NoStop}%
\bibitem [{\citenamefont {Rodrigues}\ \emph {et~al.}(2016)\citenamefont
  {Rodrigues} \emph {et~al.}}]{MINERvA:2015ydy}%
  \BibitemOpen
  \bibfield  {author} {\bibinfo {author} {\bibfnamefont {P.~A.}\ \bibnamefont
  {Rodrigues}} \emph {et~al.} (\bibinfo {collaboration} {MINERvA}),\ }\href
  {\doibase 10.1103/PhysRevLett.116.071802} {\bibfield  {journal} {\bibinfo
  {journal} {Phys. Rev. Lett.}\ }\textbf {\bibinfo {volume} {116}},\ \bibinfo
  {pages} {071802} (\bibinfo {year} {2016})},\ \bibinfo {note} {[Addendum:
  Phys.Rev.Lett. 121, 209902 (2018)]},\ \Eprint
  {http://arxiv.org/abs/1511.05944} {arXiv:1511.05944 [hep-ex]} \BibitemShut
  {NoStop}%
\bibitem [{\citenamefont {Acero}\ \emph {et~al.}(2020)\citenamefont {Acero}
  \emph {et~al.}}]{NOvA:2020rbg}%
  \BibitemOpen
  \bibfield  {author} {\bibinfo {author} {\bibfnamefont {M.~A.}\ \bibnamefont
  {Acero}} \emph {et~al.} (\bibinfo {collaboration} {NOvA}),\ }\href {\doibase
  10.1140/epjc/s10052-020-08577-5} {\bibfield  {journal} {\bibinfo  {journal}
  {Eur. Phys. J. C}\ }\textbf {\bibinfo {volume} {80}},\ \bibinfo {pages}
  {1119} (\bibinfo {year} {2020})},\ \Eprint {http://arxiv.org/abs/2006.08727}
  {arXiv:2006.08727 [hep-ex]} \BibitemShut {NoStop}%
\bibitem [{\citenamefont {Martini}\ \emph {et~al.}(2010)\citenamefont
  {Martini}, \citenamefont {Ericson}, \citenamefont {Chanfray},\ and\
  \citenamefont {Marteau}}]{Martini:2010ex}%
  \BibitemOpen
  \bibfield  {author} {\bibinfo {author} {\bibfnamefont {M.}~\bibnamefont
  {Martini}}, \bibinfo {author} {\bibfnamefont {M.}~\bibnamefont {Ericson}},
  \bibinfo {author} {\bibfnamefont {G.}~\bibnamefont {Chanfray}}, \ and\
  \bibinfo {author} {\bibfnamefont {J.}~\bibnamefont {Marteau}},\ }\href
  {\doibase 10.1103/PhysRevC.81.045502} {\bibfield  {journal} {\bibinfo
  {journal} {Phys. Rev. C}\ }\textbf {\bibinfo {volume} {81}},\ \bibinfo
  {pages} {045502} (\bibinfo {year} {2010})},\ \Eprint
  {http://arxiv.org/abs/1002.4538} {arXiv:1002.4538 [hep-ph]} \BibitemShut
  {NoStop}%
\bibitem [{\citenamefont {Martini}\ \emph {et~al.}(2011)\citenamefont
  {Martini}, \citenamefont {Ericson},\ and\ \citenamefont
  {Chanfray}}]{Martini:2011wp}%
  \BibitemOpen
  \bibfield  {author} {\bibinfo {author} {\bibfnamefont {M.}~\bibnamefont
  {Martini}}, \bibinfo {author} {\bibfnamefont {M.}~\bibnamefont {Ericson}}, \
  and\ \bibinfo {author} {\bibfnamefont {G.}~\bibnamefont {Chanfray}},\ }\href
  {\doibase 10.1103/PhysRevC.84.055502} {\bibfield  {journal} {\bibinfo
  {journal} {Phys. Rev. C}\ }\textbf {\bibinfo {volume} {84}},\ \bibinfo
  {pages} {055502} (\bibinfo {year} {2011})},\ \Eprint
  {http://arxiv.org/abs/1110.0221} {arXiv:1110.0221 [nucl-th]} \BibitemShut
  {NoStop}%
\bibitem [{\citenamefont {Nieves}\ \emph {et~al.}(2011)\citenamefont {Nieves},
  \citenamefont {Ruiz~Simo},\ and\ \citenamefont
  {Vicente~Vacas}}]{Nieves:2011pp}%
  \BibitemOpen
  \bibfield  {author} {\bibinfo {author} {\bibfnamefont {J.}~\bibnamefont
  {Nieves}}, \bibinfo {author} {\bibfnamefont {I.}~\bibnamefont {Ruiz~Simo}}, \
  and\ \bibinfo {author} {\bibfnamefont {M.~J.}\ \bibnamefont
  {Vicente~Vacas}},\ }\href {\doibase 10.1103/PhysRevC.83.045501} {\bibfield
  {journal} {\bibinfo  {journal} {Phys. Rev. C}\ }\textbf {\bibinfo {volume}
  {83}},\ \bibinfo {pages} {045501} (\bibinfo {year} {2011})},\ \Eprint
  {http://arxiv.org/abs/1102.2777} {arXiv:1102.2777 [hep-ph]} \BibitemShut
  {NoStop}%
\bibitem [{\citenamefont {Nieves}\ \emph {et~al.}(2012)\citenamefont {Nieves},
  \citenamefont {Ruiz~Simo},\ and\ \citenamefont
  {Vicente~Vacas}}]{Nieves:2011yp}%
  \BibitemOpen
  \bibfield  {author} {\bibinfo {author} {\bibfnamefont {J.}~\bibnamefont
  {Nieves}}, \bibinfo {author} {\bibfnamefont {I.}~\bibnamefont {Ruiz~Simo}}, \
  and\ \bibinfo {author} {\bibfnamefont {M.~J.}\ \bibnamefont
  {Vicente~Vacas}},\ }\href {\doibase 10.1016/j.physletb.2011.11.061}
  {\bibfield  {journal} {\bibinfo  {journal} {Phys. Lett. B}\ }\textbf
  {\bibinfo {volume} {707}},\ \bibinfo {pages} {72} (\bibinfo {year} {2012})},\
  \Eprint {http://arxiv.org/abs/1106.5374} {arXiv:1106.5374 [hep-ph]}
  \BibitemShut {NoStop}%
\bibitem [{\citenamefont {Megias}\ \emph {et~al.}(2015)\citenamefont {Megias}
  \emph {et~al.}}]{Megias:2014qva}%
  \BibitemOpen
  \bibfield  {author} {\bibinfo {author} {\bibfnamefont {G.~D.}\ \bibnamefont
  {Megias}} \emph {et~al.},\ }\href {\doibase 10.1103/PhysRevD.91.073004}
  {\bibfield  {journal} {\bibinfo  {journal} {Phys. Rev. D}\ }\textbf {\bibinfo
  {volume} {91}},\ \bibinfo {pages} {073004} (\bibinfo {year} {2015})},\
  \Eprint {http://arxiv.org/abs/1412.1822} {arXiv:1412.1822 [nucl-th]}
  \BibitemShut {NoStop}%
\bibitem [{\citenamefont {Gonzal\'ez-Jim\'enez}\ \emph
  {et~al.}(2014)\citenamefont {Gonzal\'ez-Jim\'enez}, \citenamefont {Megias},
  \citenamefont {Barbaro}, \citenamefont {Caballero},\ and\ \citenamefont
  {Donnelly}}]{Gonzalez-Jimenez:2014eqa}%
  \BibitemOpen
  \bibfield  {author} {\bibinfo {author} {\bibfnamefont {R.}~\bibnamefont
  {Gonzal\'ez-Jim\'enez}}, \bibinfo {author} {\bibfnamefont {G.~D.}\
  \bibnamefont {Megias}}, \bibinfo {author} {\bibfnamefont {M.~B.}\
  \bibnamefont {Barbaro}}, \bibinfo {author} {\bibfnamefont {J.~A.}\
  \bibnamefont {Caballero}}, \ and\ \bibinfo {author} {\bibfnamefont {T.~W.}\
  \bibnamefont {Donnelly}},\ }\href {\doibase 10.1103/PhysRevC.90.035501}
  {\bibfield  {journal} {\bibinfo  {journal} {Phys. Rev. C}\ }\textbf {\bibinfo
  {volume} {90}},\ \bibinfo {pages} {035501} (\bibinfo {year} {2014})},\
  \Eprint {http://arxiv.org/abs/1407.8346} {arXiv:1407.8346 [nucl-th]}
  \BibitemShut {NoStop}%
\bibitem [{\citenamefont {Pandey}\ \emph {et~al.}(2015)\citenamefont {Pandey},
  \citenamefont {Jachowicz}, \citenamefont {Van~Cuyck}, \citenamefont
  {Ryckebusch},\ and\ \citenamefont {Martini}}]{Pandey:2014tza}%
  \BibitemOpen
  \bibfield  {author} {\bibinfo {author} {\bibfnamefont {V.}~\bibnamefont
  {Pandey}}, \bibinfo {author} {\bibfnamefont {N.}~\bibnamefont {Jachowicz}},
  \bibinfo {author} {\bibfnamefont {T.}~\bibnamefont {Van~Cuyck}}, \bibinfo
  {author} {\bibfnamefont {J.}~\bibnamefont {Ryckebusch}}, \ and\ \bibinfo
  {author} {\bibfnamefont {M.}~\bibnamefont {Martini}},\ }\href {\doibase
  10.1103/PhysRevC.92.024606} {\bibfield  {journal} {\bibinfo  {journal} {Phys.
  Rev. C}\ }\textbf {\bibinfo {volume} {92}},\ \bibinfo {pages} {024606}
  (\bibinfo {year} {2015})},\ \Eprint {http://arxiv.org/abs/1412.4624}
  {arXiv:1412.4624 [nucl-th]} \BibitemShut {NoStop}%
\bibitem [{\citenamefont {Rocco}\ \emph
  {et~al.}(2019{\natexlab{b}})\citenamefont {Rocco}, \citenamefont {Barbieri},
  \citenamefont {Benhar}, \citenamefont {De~Pace},\ and\ \citenamefont
  {Lovato}}]{Rocco:2018mwt}%
  \BibitemOpen
  \bibfield  {author} {\bibinfo {author} {\bibfnamefont {N.}~\bibnamefont
  {Rocco}}, \bibinfo {author} {\bibfnamefont {C.}~\bibnamefont {Barbieri}},
  \bibinfo {author} {\bibfnamefont {O.}~\bibnamefont {Benhar}}, \bibinfo
  {author} {\bibfnamefont {A.}~\bibnamefont {De~Pace}}, \ and\ \bibinfo
  {author} {\bibfnamefont {A.}~\bibnamefont {Lovato}},\ }\href {\doibase
  10.1103/PhysRevC.99.025502} {\bibfield  {journal} {\bibinfo  {journal} {Phys.
  Rev. C}\ }\textbf {\bibinfo {volume} {99}},\ \bibinfo {pages} {025502}
  (\bibinfo {year} {2019}{\natexlab{b}})},\ \Eprint
  {http://arxiv.org/abs/1810.07647} {arXiv:1810.07647 [nucl-th]} \BibitemShut
  {NoStop}%
\bibitem [{\citenamefont {Ruiz~Simo}\ \emph {et~al.}(2017)\citenamefont
  {Ruiz~Simo}, \citenamefont {Amaro}, \citenamefont {Barbaro}, \citenamefont
  {De~Pace}, \citenamefont {Caballero},\ and\ \citenamefont
  {Donnelly}}]{RuizSimo:2016rtu}%
  \BibitemOpen
  \bibfield  {author} {\bibinfo {author} {\bibfnamefont {I.}~\bibnamefont
  {Ruiz~Simo}}, \bibinfo {author} {\bibfnamefont {J.~E.}\ \bibnamefont
  {Amaro}}, \bibinfo {author} {\bibfnamefont {M.~B.}\ \bibnamefont {Barbaro}},
  \bibinfo {author} {\bibfnamefont {A.}~\bibnamefont {De~Pace}}, \bibinfo
  {author} {\bibfnamefont {J.~A.}\ \bibnamefont {Caballero}}, \ and\ \bibinfo
  {author} {\bibfnamefont {T.~W.}\ \bibnamefont {Donnelly}},\ }\href {\doibase
  10.1088/1361-6471/aa6a06} {\bibfield  {journal} {\bibinfo  {journal} {J.
  Phys. G}\ }\textbf {\bibinfo {volume} {44}},\ \bibinfo {pages} {065105}
  (\bibinfo {year} {2017})},\ \Eprint {http://arxiv.org/abs/1604.08423}
  {arXiv:1604.08423 [nucl-th]} \BibitemShut {NoStop}%
\bibitem [{\citenamefont {Hernandez}\ \emph {et~al.}(2007)\citenamefont
  {Hernandez}, \citenamefont {Nieves},\ and\ \citenamefont
  {Valverde}}]{Hernandez:2007qq}%
  \BibitemOpen
  \bibfield  {author} {\bibinfo {author} {\bibfnamefont {E.}~\bibnamefont
  {Hernandez}}, \bibinfo {author} {\bibfnamefont {J.}~\bibnamefont {Nieves}}, \
  and\ \bibinfo {author} {\bibfnamefont {M.}~\bibnamefont {Valverde}},\ }\href
  {\doibase 10.1103/PhysRevD.76.033005} {\bibfield  {journal} {\bibinfo
  {journal} {Phys. Rev. D}\ }\textbf {\bibinfo {volume} {76}},\ \bibinfo
  {pages} {033005} (\bibinfo {year} {2007})},\ \Eprint
  {http://arxiv.org/abs/hep-ph/0701149} {arXiv:hep-ph/0701149} \BibitemShut
  {NoStop}%
\bibitem [{\citenamefont {Lee}(1983)}]{Lee:1983xu}%
  \BibitemOpen
  \bibfield  {author} {\bibinfo {author} {\bibfnamefont {T.~s.~h.}\
  \bibnamefont {Lee}},\ }\href {\doibase 10.1103/PhysRevLett.50.1571}
  {\bibfield  {journal} {\bibinfo  {journal} {Phys. Rev. Lett.}\ }\textbf
  {\bibinfo {volume} {50}},\ \bibinfo {pages} {1571} (\bibinfo {year}
  {1983})}\BibitemShut {NoStop}%
\bibitem [{\citenamefont {Lee}(1984)}]{Lee:1984us}%
  \BibitemOpen
  \bibfield  {author} {\bibinfo {author} {\bibfnamefont {T.~S.~H.}\
  \bibnamefont {Lee}},\ }\href {\doibase 10.1103/PhysRevC.29.195} {\bibfield
  {journal} {\bibinfo  {journal} {Phys. Rev. C}\ }\textbf {\bibinfo {volume}
  {29}},\ \bibinfo {pages} {195} (\bibinfo {year} {1984})}\BibitemShut
  {NoStop}%
\bibitem [{\citenamefont {Lee}\ and\ \citenamefont
  {Matsuyama}(1985)}]{Lee:1985jq}%
  \BibitemOpen
  \bibfield  {author} {\bibinfo {author} {\bibfnamefont {T.~S.~H.}\
  \bibnamefont {Lee}}\ and\ \bibinfo {author} {\bibfnamefont {A.}~\bibnamefont
  {Matsuyama}},\ }\href {\doibase 10.1103/PhysRevC.32.516} {\bibfield
  {journal} {\bibinfo  {journal} {Phys. Rev. C}\ }\textbf {\bibinfo {volume}
  {32}},\ \bibinfo {pages} {516} (\bibinfo {year} {1985})}\BibitemShut
  {NoStop}%
\bibitem [{\citenamefont {Lee}\ and\ \citenamefont
  {Matsuyama}(1987)}]{Lee:1987hd}%
  \BibitemOpen
  \bibfield  {author} {\bibinfo {author} {\bibfnamefont {T.~S.~H.}\
  \bibnamefont {Lee}}\ and\ \bibinfo {author} {\bibfnamefont {A.}~\bibnamefont
  {Matsuyama}},\ }\href {\doibase 10.1103/PhysRevC.36.1459} {\bibfield
  {journal} {\bibinfo  {journal} {Phys. Rev. C}\ }\textbf {\bibinfo {volume}
  {36}},\ \bibinfo {pages} {1459} (\bibinfo {year} {1987})}\BibitemShut
  {NoStop}%
\bibitem [{\citenamefont {Amaro}\ \emph {et~al.}(1998)\citenamefont {Amaro},
  \citenamefont {Barbaro}, \citenamefont {Caballero}, \citenamefont
  {Donnelly},\ and\ \citenamefont {Molinari}}]{Amaro:1998ta}%
  \BibitemOpen
  \bibfield  {author} {\bibinfo {author} {\bibfnamefont {J.~E.}\ \bibnamefont
  {Amaro}}, \bibinfo {author} {\bibfnamefont {M.~B.}\ \bibnamefont {Barbaro}},
  \bibinfo {author} {\bibfnamefont {J.~A.}\ \bibnamefont {Caballero}}, \bibinfo
  {author} {\bibfnamefont {T.~W.}\ \bibnamefont {Donnelly}}, \ and\ \bibinfo
  {author} {\bibfnamefont {A.}~\bibnamefont {Molinari}},\ }\href {\doibase
  10.1016/S0375-9474(98)00569-7} {\bibfield  {journal} {\bibinfo  {journal}
  {Nucl. Phys. A}\ }\textbf {\bibinfo {volume} {643}},\ \bibinfo {pages} {349}
  (\bibinfo {year} {1998})},\ \Eprint {http://arxiv.org/abs/nucl-th/9806014}
  {arXiv:nucl-th/9806014} \BibitemShut {NoStop}%
\bibitem [{\citenamefont {Amaro}\ \emph
  {et~al.}(2002{\natexlab{b}})\citenamefont {Amaro}, \citenamefont {Barbaro},
  \citenamefont {Caballero}, \citenamefont {Donnelly},\ and\ \citenamefont
  {Molinari}}]{Amaro:2001xz}%
  \BibitemOpen
  \bibfield  {author} {\bibinfo {author} {\bibfnamefont {J.~E.}\ \bibnamefont
  {Amaro}}, \bibinfo {author} {\bibfnamefont {M.~B.}\ \bibnamefont {Barbaro}},
  \bibinfo {author} {\bibfnamefont {J.~A.}\ \bibnamefont {Caballero}}, \bibinfo
  {author} {\bibfnamefont {T.~W.}\ \bibnamefont {Donnelly}}, \ and\ \bibinfo
  {author} {\bibfnamefont {A.}~\bibnamefont {Molinari}},\ }\href {\doibase
  10.1016/S0375-9474(01)01253-2} {\bibfield  {journal} {\bibinfo  {journal}
  {Nucl. Phys. A}\ }\textbf {\bibinfo {volume} {697}},\ \bibinfo {pages} {388}
  (\bibinfo {year} {2002}{\natexlab{b}})},\ \Eprint
  {http://arxiv.org/abs/nucl-th/0106035} {arXiv:nucl-th/0106035} \BibitemShut
  {NoStop}%
\bibitem [{\citenamefont {Jourdan}(1996)}]{Jourdan:1996np}%
  \BibitemOpen
  \bibfield  {author} {\bibinfo {author} {\bibfnamefont {J.}~\bibnamefont
  {Jourdan}},\ }\href {\doibase 10.1016/0375-9474(96)00143-1} {\bibfield
  {journal} {\bibinfo  {journal} {Nucl. Phys. A}\ }\textbf {\bibinfo {volume}
  {603}},\ \bibinfo {pages} {117} (\bibinfo {year} {1996})}\BibitemShut
  {NoStop}%
\bibitem [{\citenamefont {Amaro}\ \emph {et~al.}(2003)\citenamefont {Amaro},
  \citenamefont {Barbaro}, \citenamefont {Caballero}, \citenamefont
  {Donnelly},\ and\ \citenamefont {Molinari}}]{amaro2003delta}%
  \BibitemOpen
  \bibfield  {author} {\bibinfo {author} {\bibfnamefont {J.}~\bibnamefont
  {Amaro}}, \bibinfo {author} {\bibfnamefont {M.~B.}\ \bibnamefont {Barbaro}},
  \bibinfo {author} {\bibfnamefont {J.}~\bibnamefont {Caballero}}, \bibinfo
  {author} {\bibfnamefont {T.}~\bibnamefont {Donnelly}}, \ and\ \bibinfo
  {author} {\bibfnamefont {A.}~\bibnamefont {Molinari}},\ }\href@noop {}
  {\bibfield  {journal} {\bibinfo  {journal} {Nuclear Physics A}\ }\textbf
  {\bibinfo {volume} {723}},\ \bibinfo {pages} {181} (\bibinfo {year}
  {2003})}\BibitemShut {NoStop}%
\bibitem [{\citenamefont {Korover}\ \emph {et~al.}(2021)\citenamefont {Korover}
  \emph {et~al.}}]{CLAS:2020rue}%
  \BibitemOpen
  \bibfield  {author} {\bibinfo {author} {\bibfnamefont {I.}~\bibnamefont
  {Korover}} \emph {et~al.} (\bibinfo {collaboration} {CLAS}),\ }\href
  {\doibase 10.1016/j.physletb.2021.136523} {\bibfield  {journal} {\bibinfo
  {journal} {Phys. Lett. B}\ }\textbf {\bibinfo {volume} {820}},\ \bibinfo
  {pages} {136523} (\bibinfo {year} {2021})},\ \Eprint
  {http://arxiv.org/abs/2004.07304} {arXiv:2004.07304 [nucl-ex]} \BibitemShut
  {NoStop}%
\bibitem [{\citenamefont {Duer}\ \emph {et~al.}(2019)\citenamefont {Duer} \emph
  {et~al.}}]{CLAS:2018xvc}%
  \BibitemOpen
  \bibfield  {author} {\bibinfo {author} {\bibfnamefont {M.}~\bibnamefont
  {Duer}} \emph {et~al.} (\bibinfo {collaboration} {CLAS}),\ }\href {\doibase
  10.1103/PhysRevLett.122.172502} {\bibfield  {journal} {\bibinfo  {journal}
  {Phys. Rev. Lett.}\ }\textbf {\bibinfo {volume} {122}},\ \bibinfo {pages}
  {172502} (\bibinfo {year} {2019})},\ \Eprint
  {http://arxiv.org/abs/1810.05343} {arXiv:1810.05343 [nucl-ex]} \BibitemShut
  {NoStop}%
\bibitem [{\citenamefont {Aguilar-Arevalo}\ \emph {et~al.}(2010)\citenamefont
  {Aguilar-Arevalo} \emph {et~al.}}]{MiniBooNE:2010bsu}%
  \BibitemOpen
  \bibfield  {author} {\bibinfo {author} {\bibfnamefont {A.~A.}\ \bibnamefont
  {Aguilar-Arevalo}} \emph {et~al.} (\bibinfo {collaboration} {MiniBooNE}),\
  }\href {\doibase 10.1103/PhysRevD.81.092005} {\bibfield  {journal} {\bibinfo
  {journal} {Phys. Rev. D}\ }\textbf {\bibinfo {volume} {81}},\ \bibinfo
  {pages} {092005} (\bibinfo {year} {2010})},\ \Eprint
  {http://arxiv.org/abs/1002.2680} {arXiv:1002.2680 [hep-ex]} \BibitemShut
  {NoStop}%
\bibitem [{\citenamefont {Bodek}\ \emph {et~al.}(2008)\citenamefont {Bodek},
  \citenamefont {Avvakumov}, \citenamefont {Bradford},\ and\ \citenamefont
  {Budd}}]{Bodek:2007ym}%
  \BibitemOpen
  \bibfield  {author} {\bibinfo {author} {\bibfnamefont {A.}~\bibnamefont
  {Bodek}}, \bibinfo {author} {\bibfnamefont {S.}~\bibnamefont {Avvakumov}},
  \bibinfo {author} {\bibfnamefont {R.}~\bibnamefont {Bradford}}, \ and\
  \bibinfo {author} {\bibfnamefont {H.~S.}\ \bibnamefont {Budd}},\ }\href
  {\doibase 10.1140/epjc/s10052-007-0491-4} {\bibfield  {journal} {\bibinfo
  {journal} {Eur. Phys. J. C}\ }\textbf {\bibinfo {volume} {53}},\ \bibinfo
  {pages} {349} (\bibinfo {year} {2008})},\ \Eprint
  {http://arxiv.org/abs/0708.1946} {arXiv:0708.1946 [hep-ex]} \BibitemShut
  {NoStop}%
\bibitem [{\citenamefont {Meyer}\ \emph {et~al.}(2016)\citenamefont {Meyer},
  \citenamefont {Betancourt}, \citenamefont {Gran},\ and\ \citenamefont
  {Hill}}]{Meyer:2016oeg}%
  \BibitemOpen
  \bibfield  {author} {\bibinfo {author} {\bibfnamefont {A.~S.}\ \bibnamefont
  {Meyer}}, \bibinfo {author} {\bibfnamefont {M.}~\bibnamefont {Betancourt}},
  \bibinfo {author} {\bibfnamefont {R.}~\bibnamefont {Gran}}, \ and\ \bibinfo
  {author} {\bibfnamefont {R.~J.}\ \bibnamefont {Hill}},\ }\href {\doibase
  10.1103/PhysRevD.93.113015} {\bibfield  {journal} {\bibinfo  {journal} {Phys.
  Rev. D}\ }\textbf {\bibinfo {volume} {93}},\ \bibinfo {pages} {113015}
  (\bibinfo {year} {2016})},\ \Eprint {http://arxiv.org/abs/1603.03048}
  {arXiv:1603.03048 [hep-ph]} \BibitemShut {NoStop}%
\bibitem [{\citenamefont {Cai}\ \emph {et~al.}(2023)\citenamefont {Cai} \emph
  {et~al.}}]{MINERvA:2023avz}%
  \BibitemOpen
  \bibfield  {author} {\bibinfo {author} {\bibfnamefont {T.}~\bibnamefont
  {Cai}} \emph {et~al.} (\bibinfo {collaboration} {MINERvA}),\ }\href {\doibase
  10.1038/s41586-022-05478-3} {\bibfield  {journal} {\bibinfo  {journal}
  {Nature}\ }\textbf {\bibinfo {volume} {614}},\ \bibinfo {pages} {48}
  (\bibinfo {year} {2023})}\BibitemShut {NoStop}%
\bibitem [{\citenamefont {Meyer}\ \emph {et~al.}(2022)\citenamefont {Meyer},
  \citenamefont {Walker-Loud},\ and\ \citenamefont
  {Wilkinson}}]{Meyer:2022mix}%
  \BibitemOpen
  \bibfield  {author} {\bibinfo {author} {\bibfnamefont {A.~S.}\ \bibnamefont
  {Meyer}}, \bibinfo {author} {\bibfnamefont {A.}~\bibnamefont {Walker-Loud}},
  \ and\ \bibinfo {author} {\bibfnamefont {C.}~\bibnamefont {Wilkinson}},\
  }\href {\doibase 10.1146/annurev-nucl-010622-120608} {\bibfield  {journal}
  {\bibinfo  {journal} {Ann. Rev. Nucl. Part. Sci.}\ }\textbf {\bibinfo
  {volume} {72}},\ \bibinfo {pages} {205} (\bibinfo {year} {2022})},\ \Eprint
  {http://arxiv.org/abs/2201.01839} {arXiv:2201.01839 [hep-lat]} \BibitemShut
  {NoStop}%
\end{thebibliography}%

\end{document}